\documentclass{ws-ijmpa}
\usepackage[super,compress]{cite}
\usepackage{graphicx}
\usepackage{hyperref}
\hypersetup{colorlinks,urlcolor=black,citecolor=black,linkcolor=black,filecolor=black}

\begin{document}

\markboth{Frank Zimmermann}{``Colliders or Bust''}

%
\catchline{}{}{}{}{}
%

\title{``Colliders or Bust''\footnote{Invited Contribution to Swapan Chattopadhyay Retirement Symposium, 30 April 2021}}

\author{Frank Zimmermann}

\address{European Organization for Nuclear Research (CERN), \\
Esplanade des Particules 1, 1217 Meyrin, Switzerland\\
frank.zimmermann@cern.ch}

\maketitle

\begin{history}
\received{Day Month Year}
\revised{Day Month Year}
\end{history}

\begin{abstract}
Charged-particle colliders have proven key instruments of discovery in high-energy
physics. Pushing the frontiers of our knowledge 
ever further has relied on, and still keeps calling for, ever better performance and novel techniques. During more than four decades,  
Swapan Chattopadhyay has made numerous essential contributions to this endeavour.
Often far ahead of his time, 
he helped advance many areas of collider development, notably in the domains of stochastic cooling, the development of asymmetric B factories, the design of next- and next-next-generation of high-energy colliders, and the harnessing of energy-recovery 
for particle colliders. 
\end{abstract}

\ccode{PACS numbers:}


\section{A Collider Century}
\begin{figure}[htbp]
\centering
\includegraphics[width=0.99\linewidth]{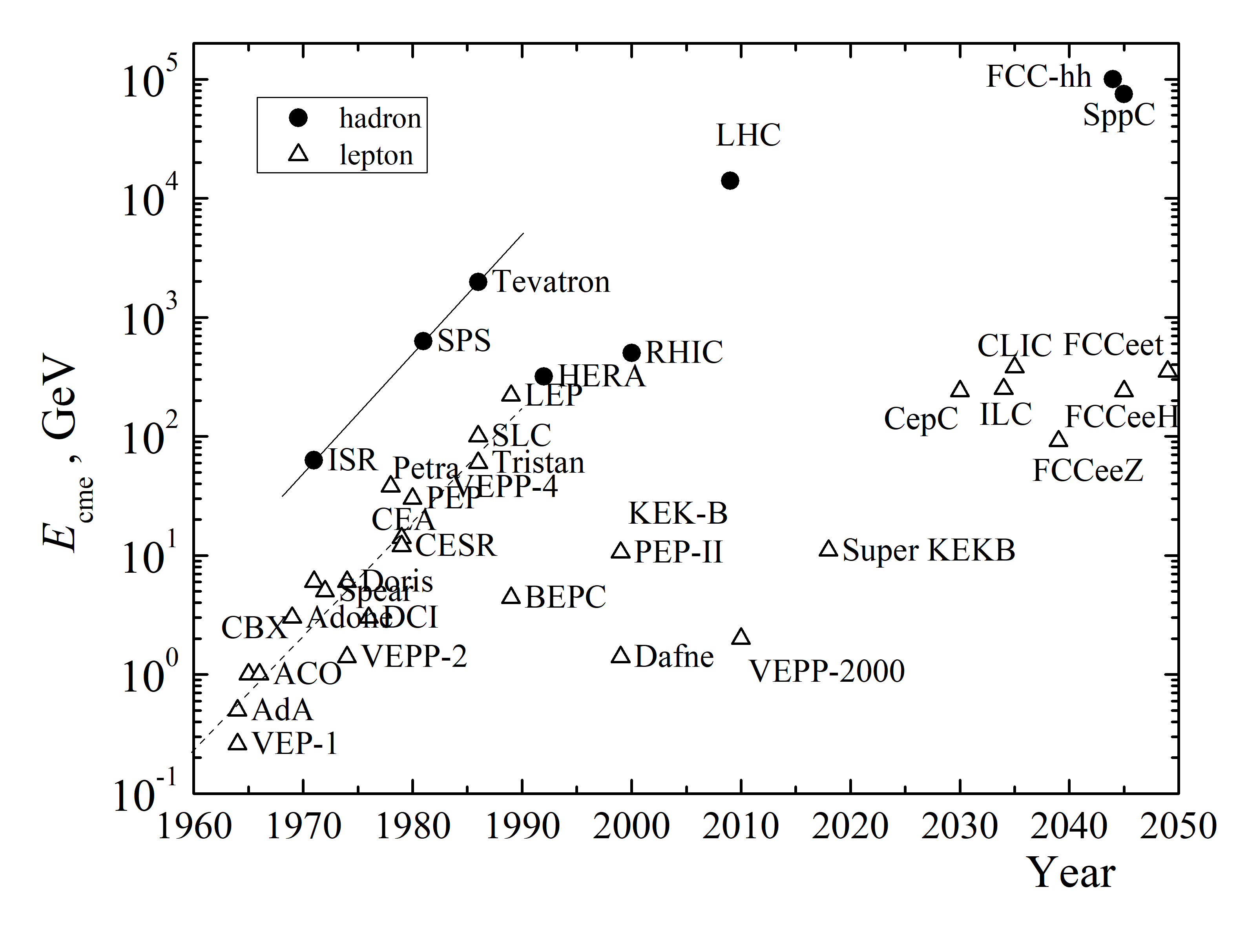}
\caption{Center of mass energy reach of particle colliders vs their start of operation. Solid and dashed lines indicate a ten-fold increase per decade for hadron (circles) and lepton (triangles) colliders (from Ref.~\cite{RevModPhys.93.015006}, adapted from Ref.~\cite{shiltsev2012}).}
\label{fig:colliders_E}
\end{figure}

Over the past 60 years, high-energy 
charged particle colliders have proven extremely efficient tools of discovery and precision physics.   
All the heavier elements of the Standard Model of particle physics were discovered (or co-discovered) at particle colliders: 
the charm quark and tau lepton at the SPEAR e$^+$e$^-$  collider,
the gluon in e$^+$e$^-$ collisions at PETRA,  the W and Z bosons in ${\rm p}\bar{\rm p}$ collisions at the S${\rm p}\bar{\rm p}$S, the top quark at 
the ${\rm p}\bar{\rm p}$ collider 
Tevatron, and the Higgs boson in proton-proton collisions 
at the Large Hadron Collider.
Also electron-proton collisions at HERA
(parton distribution functions) 
and heavy ion collisions at both 
RHIC and the LHC contributed greatly to, e.g. to our understanding of quantum chromodynamics,
the origin of nuclear spin, etc. 
Figure \ref{fig:colliders_E} presents the evolution
of the centre-of-mass energies of 
both lepton and hadron colliders starting in the 1960s,
with a tentative forecast featuring proposed future 
machines.

\begin{figure}[htbp]
\centering
\includegraphics[width=0.99\linewidth]{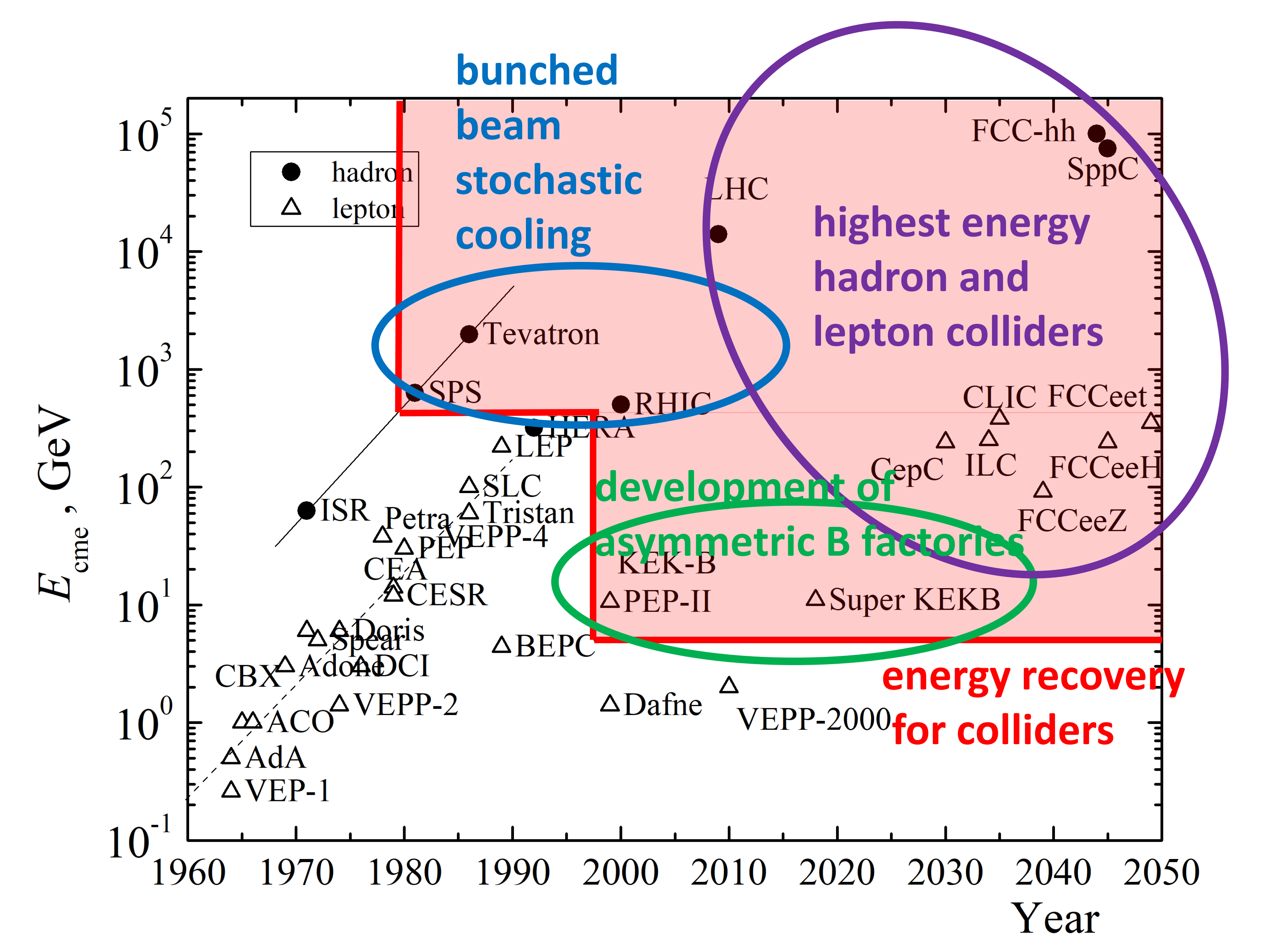}
\caption{Past, present and future colliders to which Swapan Chattopadhyay made key  contributions, indicated by ellipses and rectangular boxes   
(lines, boxes and text superimposed on Fig.~\ref{fig:colliders_E}, taken from Ref.\cite{RevModPhys.93.015006}).}
\label{fig:colliders_Swapan}
\end{figure}

It is remarkable that Swapan Chattopadhyay made key 
contributions to basically all high-energy colliders 
from the 1980s onwards, and extrapolated till 2050, 
as is illustrated in Fig.~\ref{fig:colliders_Swapan}.
Specifically: (1) he helped develop techniques of  
stochastic cooling and bunched beam stochastic 
cooling of antiprotons  and heavy ions, which proved essential for the
S${\rm p}\bar{\rm p}$S, Tevatron and RHIC;
(2) he conceived the basic concepts underpinning the   
asymmetric e$^+$e$^-$ B factories, which enabled the successes of 
PEP-II, KEKB, and SuperKEKB, with dramatically higher luminosity
than any previous machines;
(3) he contributed essential elements to the  
research and development for highest energy hadron and lepton colliders, such  
LHC, HL-LHC, FCC, CLIC, ILC, muon colliders, 
$\gamma \gamma$ colliders, and plasma-based colliders;
and (4) Swapan Chattopadhyay 
was the first to propose and promote
the use of energy recovery for the Large Hadron electron Collider (LHeC)
and variants thereof. 
The community has rarely witnessed so versatile 
an accelerator physicist,
with activities covering all kinds
of lepton and hadron colliders, including their injector chains,
along with many far-future approaches.

\section{Bunched-Beam Stochastic Cooling}
Swapan Chattopadhyay belonged to a small group of 
pioneers who developed the theory of stochastic 
cooling\cite{Chattopadhyay:1982jz,Chattopadhyay:149022}. 
In their historical review of stochastic beam cooling,
Fritz Caspers and Dieter M\"{o}hl noted\cite{Caspers:1447119}:
``Starting in 1981 S. Chattopadhyay (partly together with Bisognano)\cite{Chattopadhyay155458,Chattopadhyay:1982jz} established the theory bringing to perfection earlier treatments\cite{Herr:133439}.''

\begin{figure}[htbp]
\centering
\includegraphics[width=0.40\linewidth]{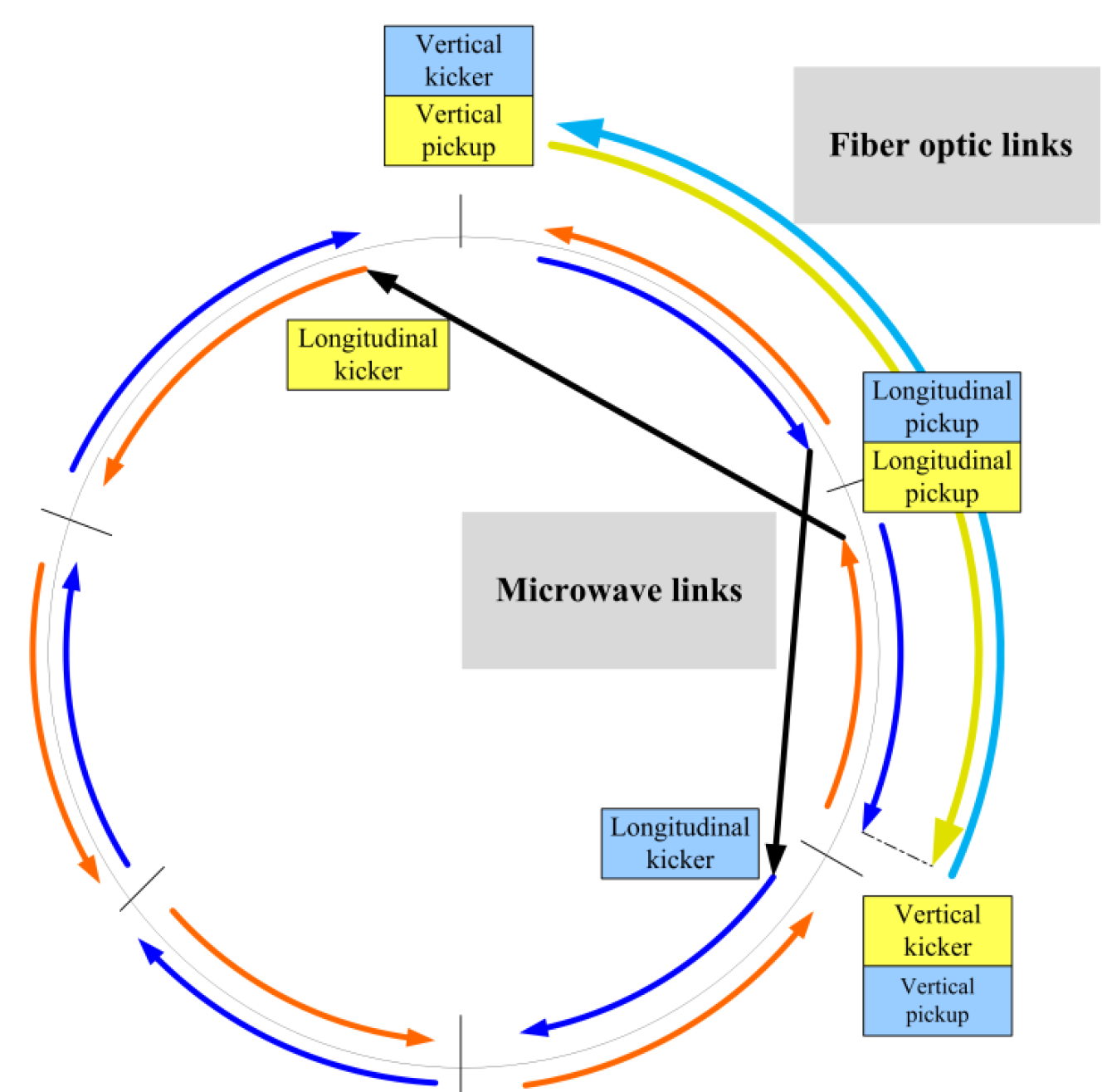}
\includegraphics[width=0.55\linewidth]{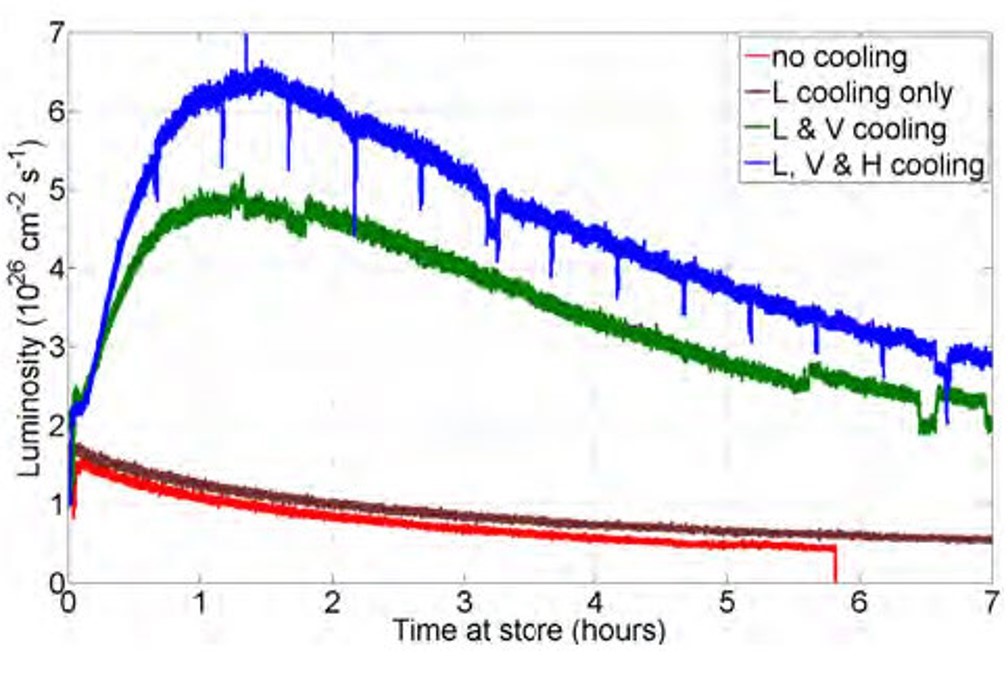}
\caption{Schematic of the bunched beam stochastic cooling system at RHIC, consisting of ``pickups'' and ``kickers'' --- and the fibre-optic and microwave links between these two; the 70 GHz microwave links for the longitudinal system are sent on received on the surface, while the fibre-optics links for transverse cooling are integrated in the tunnel
\cite{Mernick:BIW10-TUPSM085} 
(left); comparison
of RHIC luminosity as a function of time in the store,
without cooling (red), with only longitudinal cooling (brown), 2-D cooling (green), and 3-D cooling (blue) 
\protect\cite{cc2012}.}
\label{fig:RHICSC}
\end{figure}

On his own, and together with Daniel Boussard, Georges D\^{o}me, and Trevor Linnecar, Swapan Chattopadhyay 
also studied the feasibility of 
bunched beam stochastic cooling was finally achieved 
 at RHIC\cite{Blaskiewicz:2010zz}, leading to a five-fold increase in integrated RHIC luminosity, in collisions of uranium ions; see Fig.~\ref{fig:RHICSC}.

\section{Optical Stochastic Cooling}
Extending the concept of stochastic cooling to much higher frequencies and bandwidth gave rise to the idea
of optical stochastic cooling (OSC)\cite{Mikhailichenko:1993mn,Zolotorev:1994hm}, which Swapan helped to develop and attempted to demonstrate 
in the 1990s\cite{Chattopadhyay:1997rz}.  
With a few well-known 
co-authors, like Alexander Zholents and Max Zolotorev, 
he also proposed, and explored, the use of
OSC for beam halo confinement at the VLHC\cite{Zholents:2000yj}, a future highest-energy hadron collider near Fermilab, then under consideration. 
 OSC holds the promise to 
 speed up beam cooling by four orders of magnitude compared to conventional stochastic cooling,
 thanks to its much larger bandwidth. 

\begin{figure}[htbp]
\centering
\includegraphics[width=0.59\linewidth]{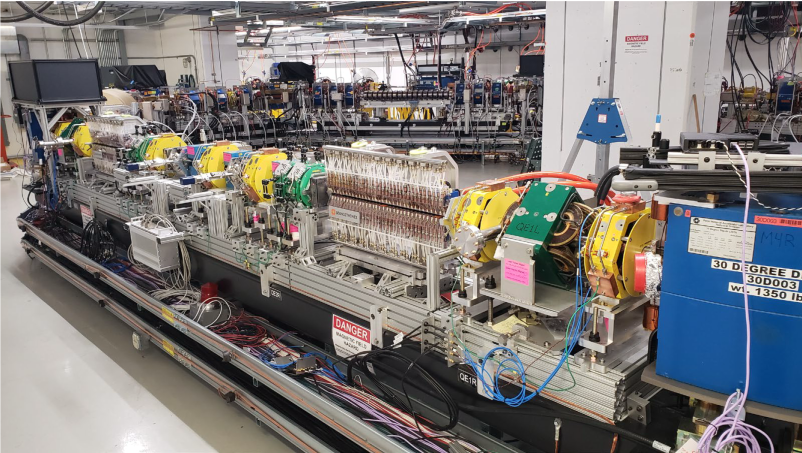}
\includegraphics[width=0.36\linewidth]{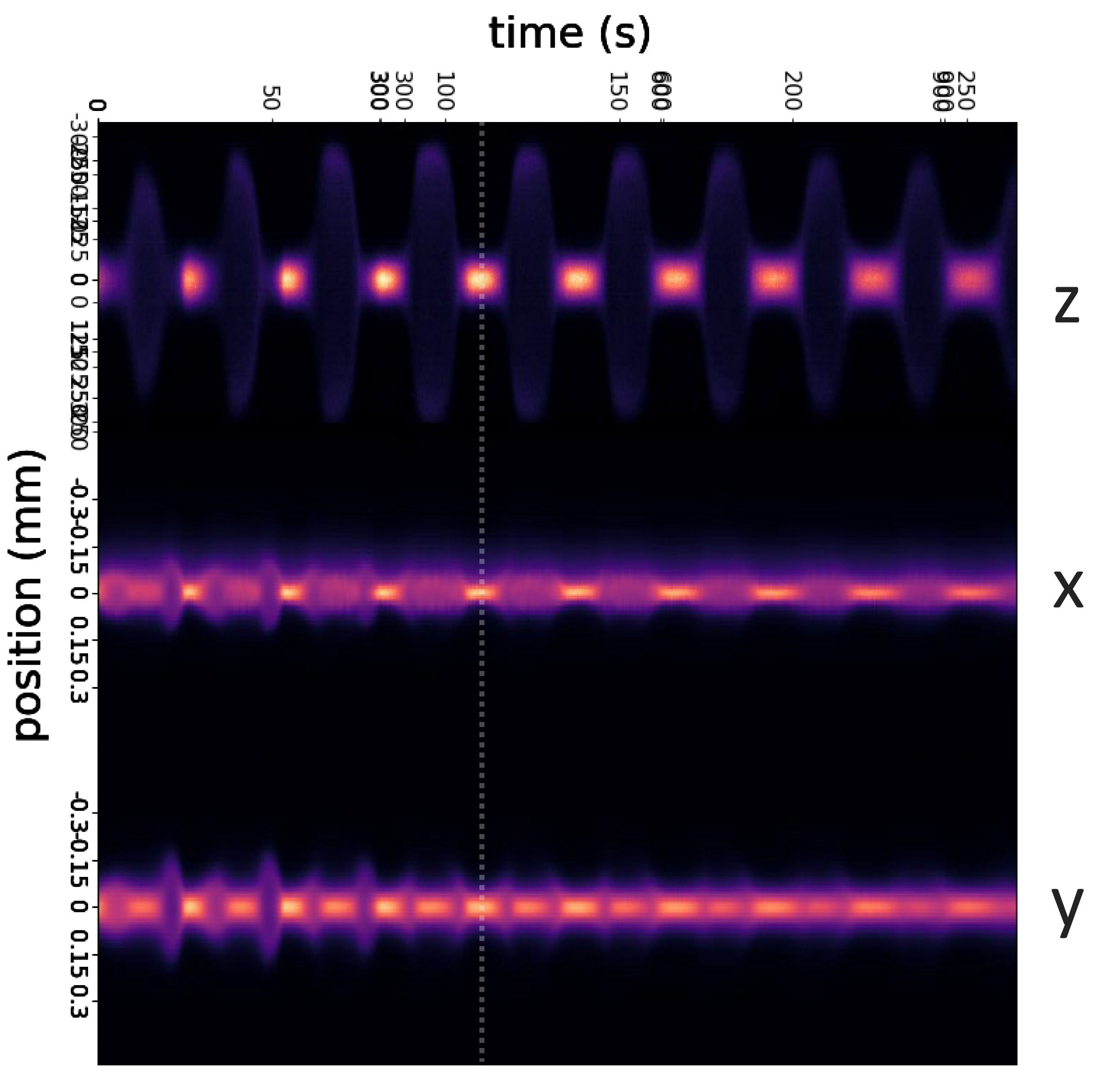}
\caption{OSC pick-up and kicker undulator installed at IOTA\cite{cool2021} (left); evidence for OSC of an electron bunch in all three dimensions\cite{cool2021} (right).}
\label{fig:IOTAOSC}
\end{figure}

It is exciting that an OSC test has now been set up 
at FNAL's IOTA facility\cite{Lebedev:2020vkh}, 
and that, here,  with an electron beam,
 3-dimensional OSC has been demonstrated for the first
 time in 2021\cite{Valishev:2021src,cool2021} 
 (see Fig.~\ref{fig:IOTAOSC}) --- 
 almost 30 years after the initial proposals. 
This breakthrough opens 
up many intriguing possibilities.

\section{Asymmetric B Factories}
\begin{figure}[htbp]
\centering
\includegraphics[width=0.95\linewidth]{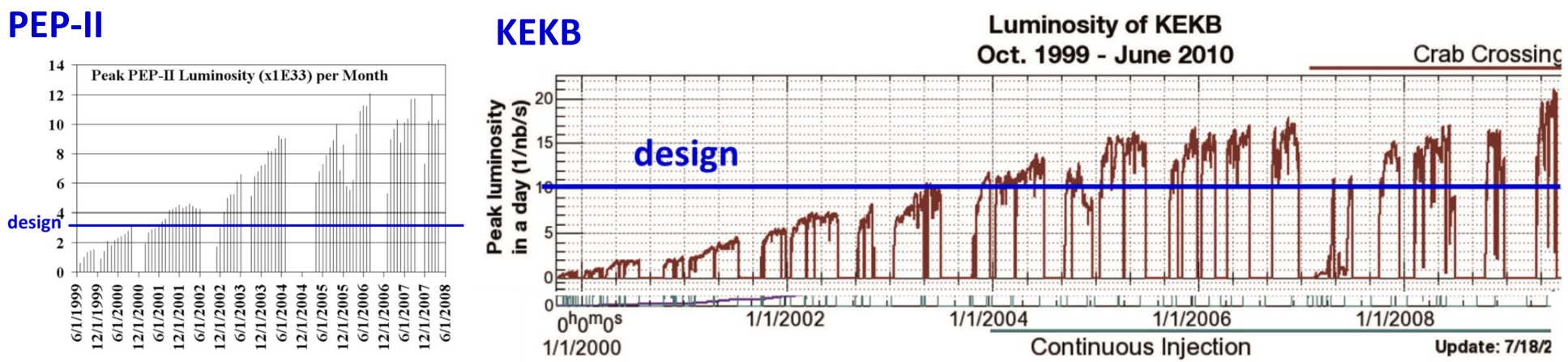}
\caption{Peak luminosity of PEP-II and KEKB over a time period of ten years compared with the respective design luminosity 
(Courtesy SLAC and KEK).}
\label{fig:AsBFactory}
\end{figure}

Swapan Chattopadhyay is one of the fathers of the highly successful asymmetric B factories, who contributed to 
developing this concept {\it ab initio}\cite{Garren:1989vh,Chattopadhyay:1989pf}.
The luminosity performance rapidly 
achieved by the two B factories PEP-II at SLAC and KEKB at KEK 
more than validated the concept  --- see Fig.~\ref{fig:AsBFactory}. 

The successes of PEP-II and KEKB 
 inspired an even more ambitious project, a
Super B factory, SuperKEKB, presently under commissioning. 
SuperKEKB aims for still another 
factor 30--40 higher luminosity than KEKB,
using a ``nanobeam'' or ``crab-waist'' collision scheme\cite{Raimondi:1182932}.

\begin{figure}[htbp]
\centering
\includegraphics[width=0.95\linewidth]{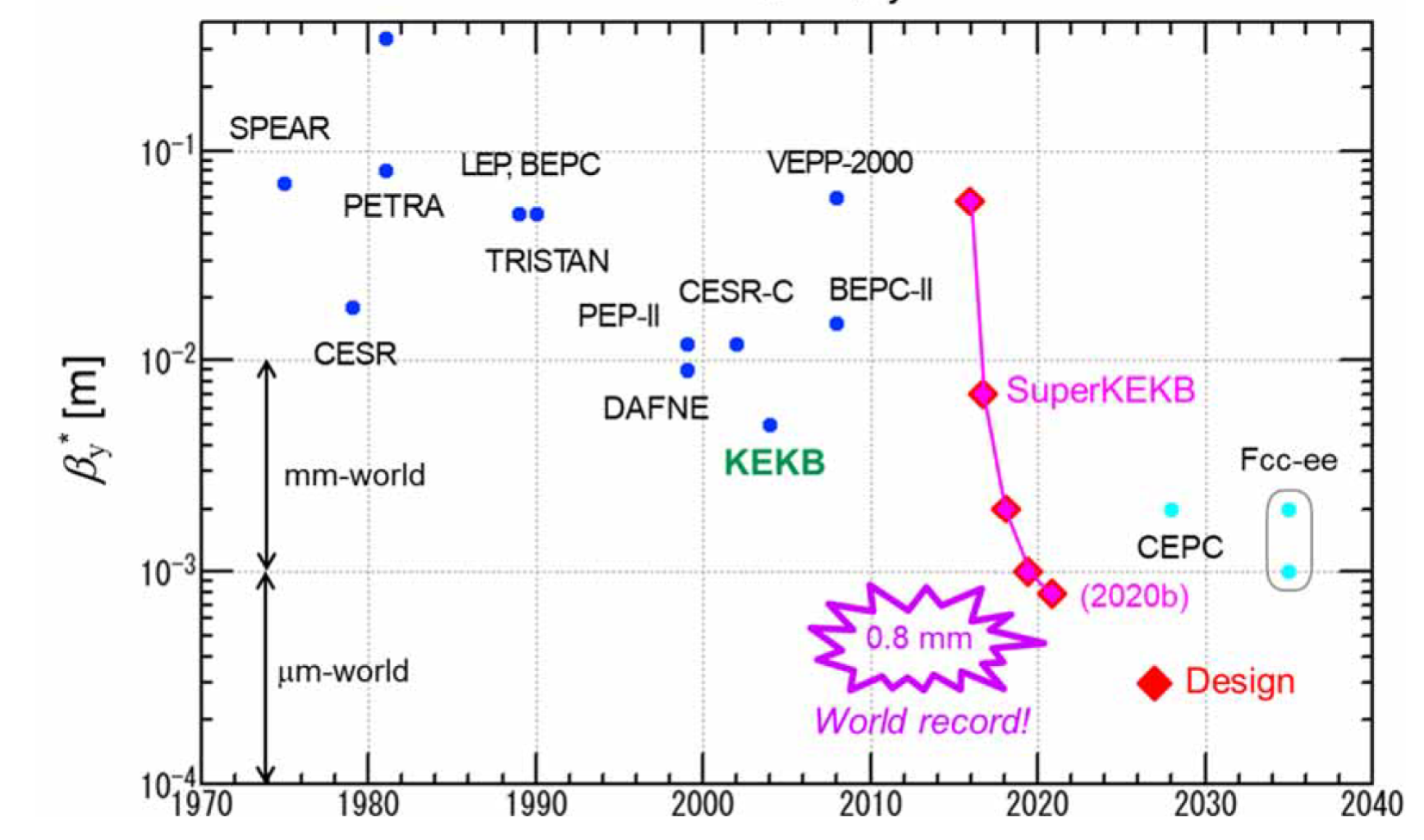}
\caption{Vertical beta function at the collision point for different
past, present and future e$^+$e$^-$ colliders versus the year 
(Courtesy K.~Shibata, M.~Tobiyama, K.~Oide, et al., KEK).}
\label{fig:betastar}
\end{figure}

In 2020, a vertical beta function at the interaction point of $\beta_{y}^{\ast}   = 0.8$~mm was achieved in both SuperKEKB  rings, 
using a ``virtual'' crab-waist collision optics first developed for the FCC-ee\cite{PhysRevAccelBeams.19.111005}. This $\beta_{y}^{\ast}$ value is a world record.
Figure~\ref{fig:betastar} puts this value in perspective by comparing with previous e$^+$e$^-$ colliders, with the design goal, and with proposed future colliders.  
On 23 December 2021, SuperKEKB also established a new world record for the peak luminosity of $3.8\times 10^{34}$~cm$^{-2}$s$^{-1}$.

\section{Highest Energy Hadron and Lepton Colliders}
It should come as no surprise that  
Swapan Chattpadhyay also contributed enormously to 
present and future highest-energy colliders, including
LHC, HL-LHC, FCC, CLIC, ILC, muon colliders, 
$\gamma\gamma$ colliders, and 
plasma-based colliders.

Swapan Chattopadhyay wrote
wrote major parts of the conceptual design report (CDR)
for the Superconducting Super Collider (SSC) from 1986\cite{Jackson:108121}. In particular, he developed an organizational structure and carried our a detailed cost estimate.

\begin{figure}[htbp]
\centering
\includegraphics[width=0.95\linewidth]{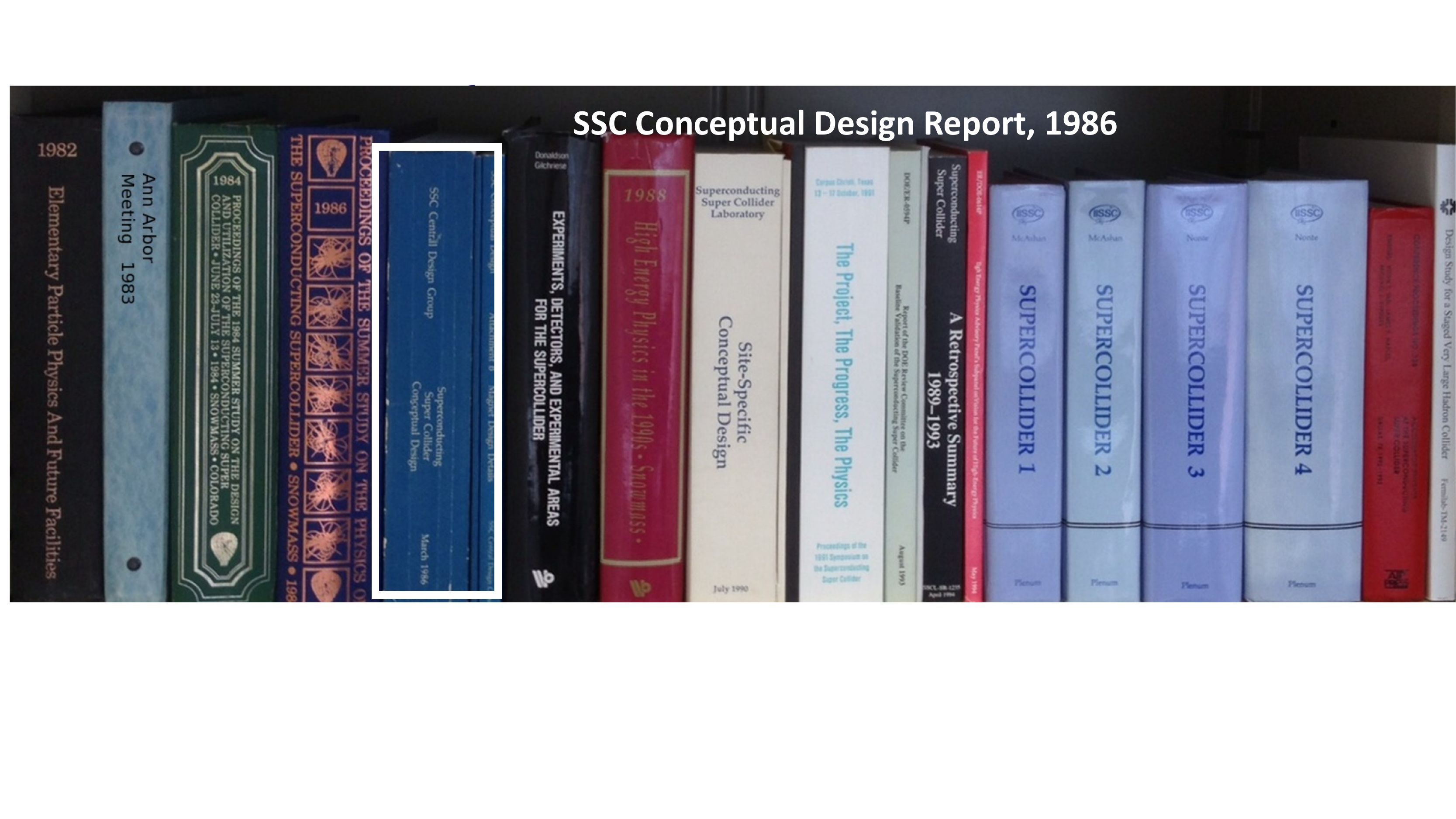}
\caption{The 1986 SSC CDR\protect\cite{Jackson:108121}  (white box), edited by J.D.~Jackson, among 
other SCC literature 
on Mike Syphers's bookshelf  (Courtesy M.~Syphers).}
\label{fig:SSCCDR}
\end{figure}
 
\begin{figure}[htbp]
\centering
\includegraphics[width=0.33\linewidth]{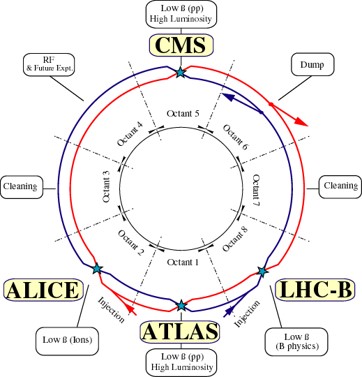}
\includegraphics[width=0.58\linewidth]{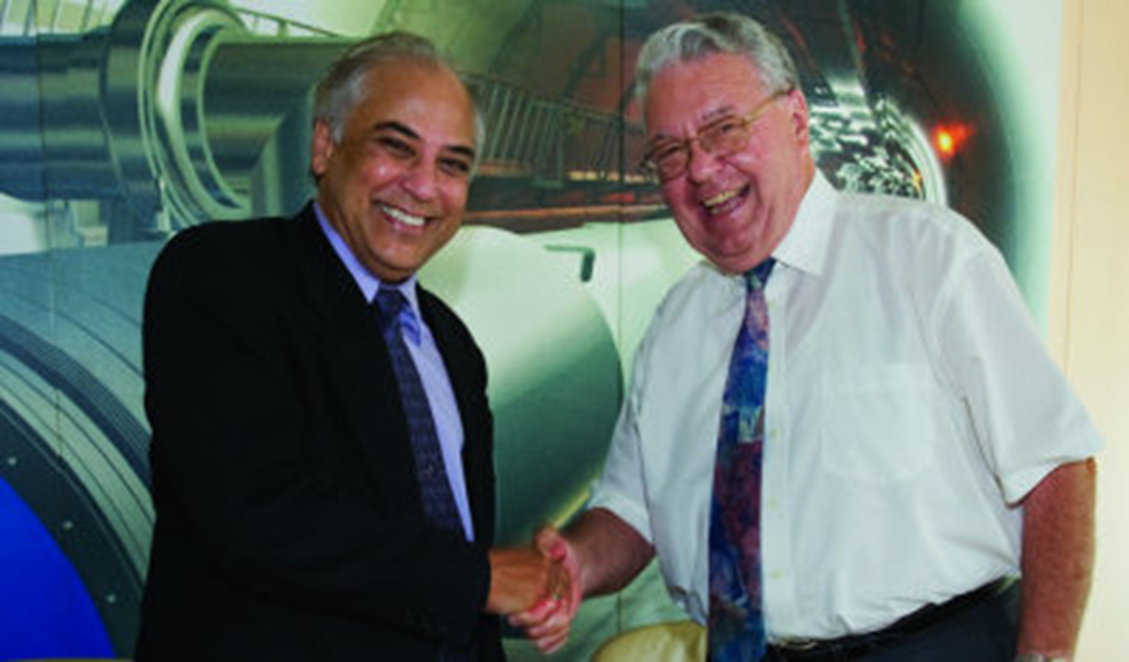}
\caption{Schematic layout of the Large Hadron Collider at CERN (left); and a photo from 2008 with CI Director Swapan Chattopadhyay and CERN Director-General Robert Aymar   
finalizing the comprehensive CI-CERN collaboration agreement 
by handshake (right); both images courtesy CERN.}
\label{fig:lhc}
\end{figure}

The present energy frontier is defined by 
the Large Hadron Collider (LHC) at CERN, 
a double ring of almost 27 km circumference (see Fig.~\ref{fig:lhc}, left picture),
which since 2010  provides proton-proton collisions,
at centre-of-mass energies that are 
approaching the design value of 14 TeV.
The LHC including its luminosity upgrade HL-LHC
is set to operate until about 2040.
Farsightedly, as director of the Cockcroft Institute (CI),  
in 2008, Swapan Chattopadhyay concluded 
a comprehensive collaboration agreement with CERN  (Fig.~\ref{fig:lhc}, right picture), cementing CI's position and establishing the foundation for the UK's subsequent leading role in the HL-LHC project.

\begin{figure}[htbp]
\centering
\includegraphics[width=0.62\linewidth]{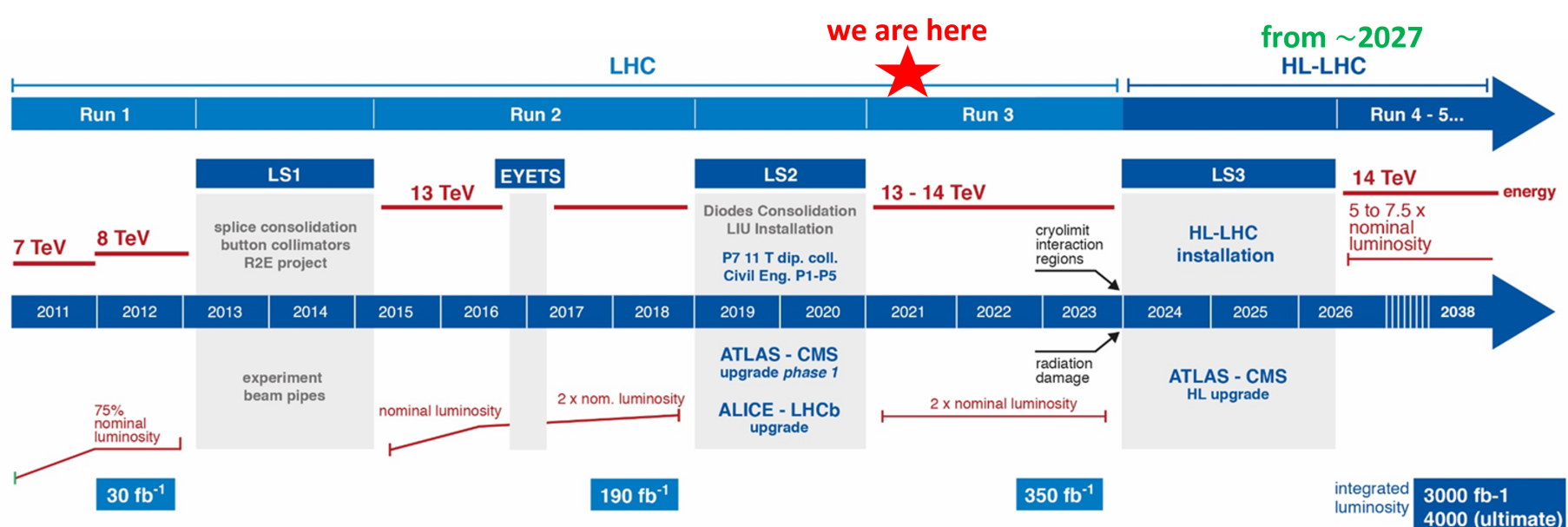}
\includegraphics[width=0.35\linewidth]{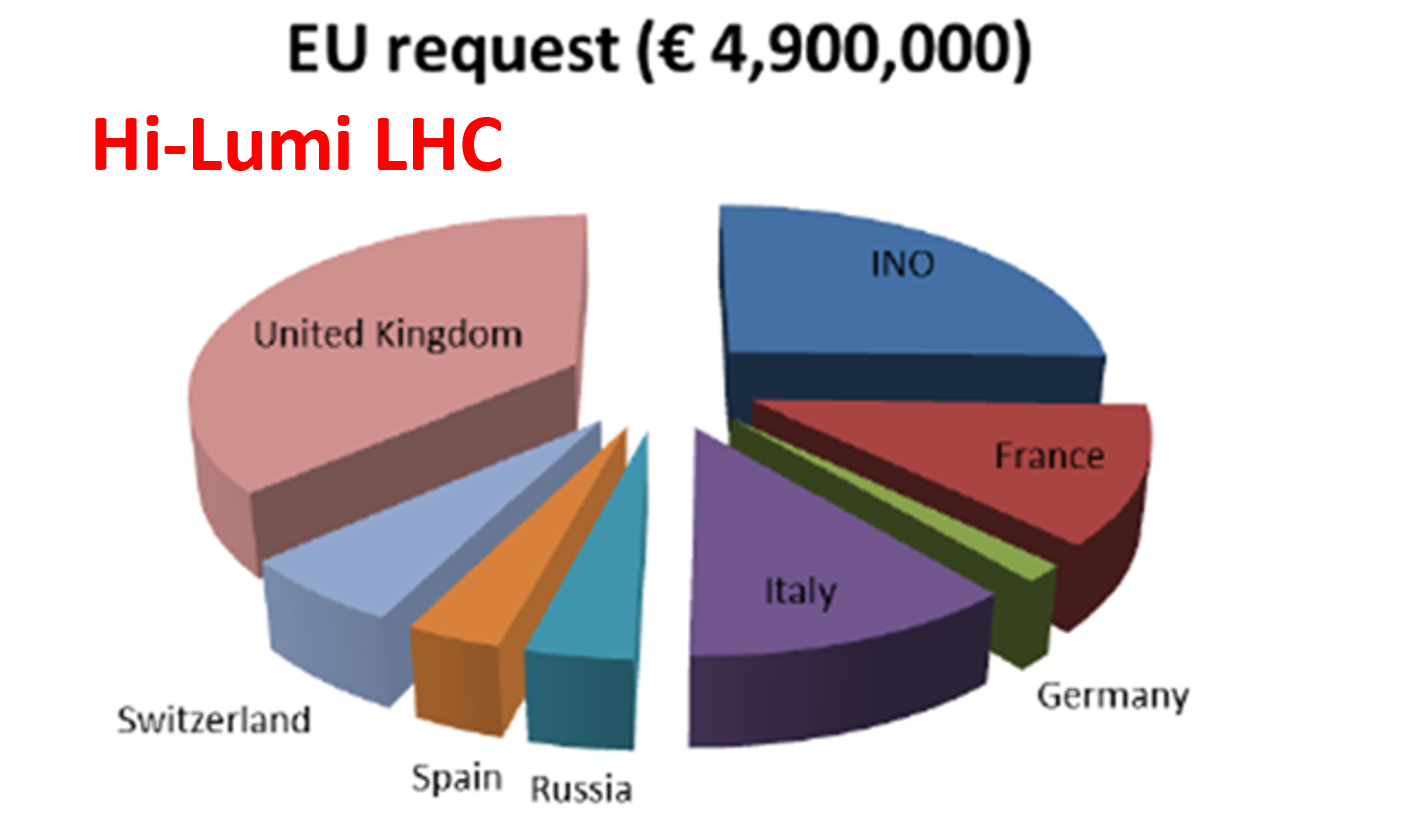}
\caption{HL-LHC time line (left) and budget distribution for the EU co-financed 
HiLumi LHC project (right); 
both images courtesy CERN.}
\label{fig:hl-lhc}
\end{figure}

Indeed, under
Swapan Chattopadhyay’s directorship, 
in particular the CI assumed a
prominent role in the LHC and Hi-Lumi LHC Upgrade, including leading the Collaboration Board, 
and carrying out pioneering developments of ``crab cavities'', other RF R\&D and multiple HL-LHC accelerator design efforts, e.g., on the machine detector interface.
Figure \ref{fig:hl-lhc} presents the HL-LHC 
time line as of 2020, and the HiLumi budget distribution, revealing that the largest share 
was attributed to the UK.

\begin{figure}[htbp]
\centering
\includegraphics[width=0.90\linewidth]{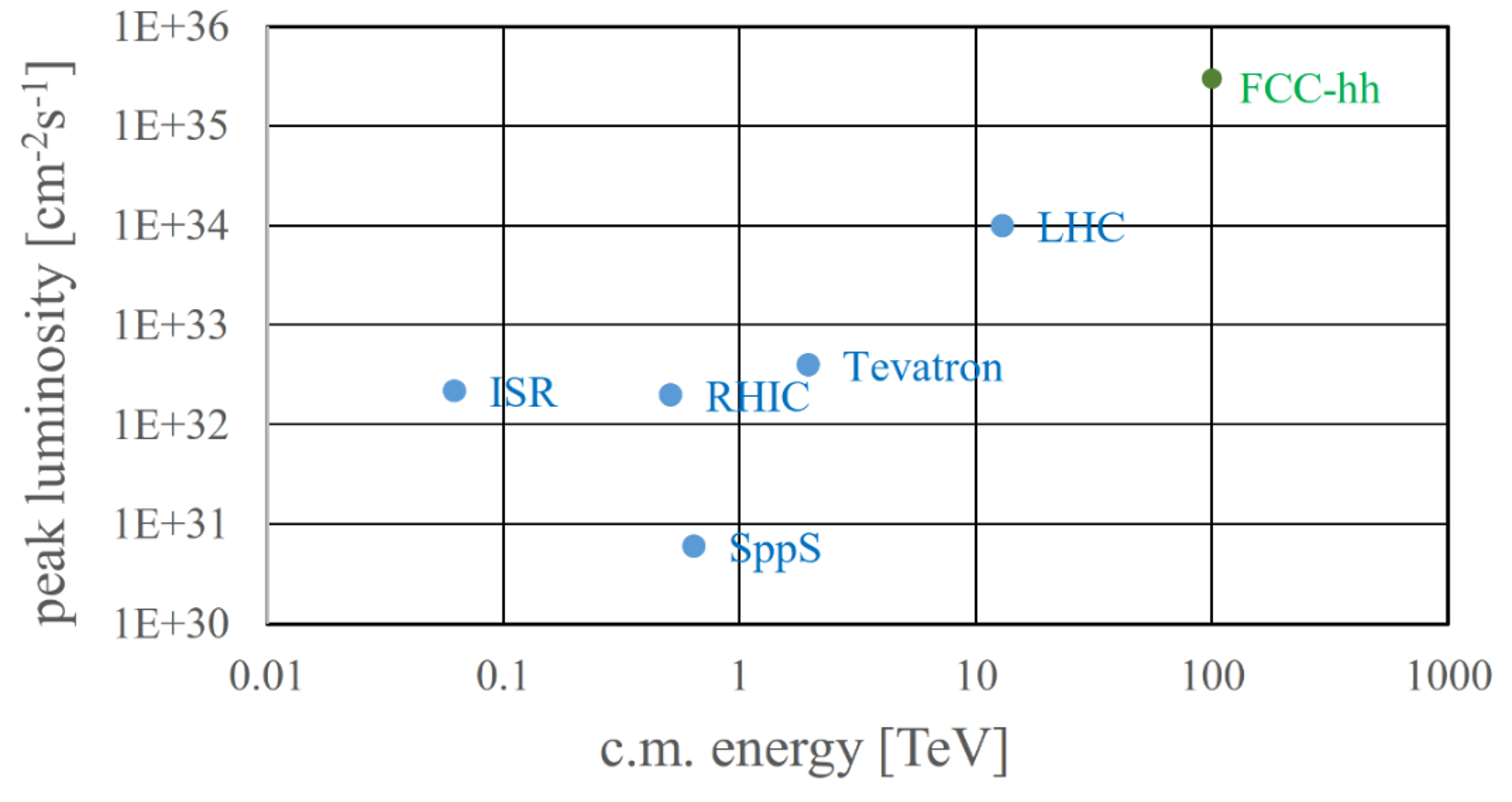}
\caption{Hadron collider peak luminosity versus centre-of-mass energy on
a double logarithmic scale; see also Ref.\cite{Benedikt:2018ofy}.}
\label{fig:hadroncoll}
\end{figure}

Following the HL-LHC the next energy frontier collider
could be the Future Circular hadron Collider FCC-hh \cite{fcchh}.
Figure \ref{fig:hadroncoll} shows the historical
path of hadron colliders in the luminosity-energy plane.
The step from the 
LHC to the FCC-hh will be as large as the step from the Tevatron to the LHC.

\begin{figure}[htb]
\begin{center}
\includegraphics[width=0.48\linewidth]{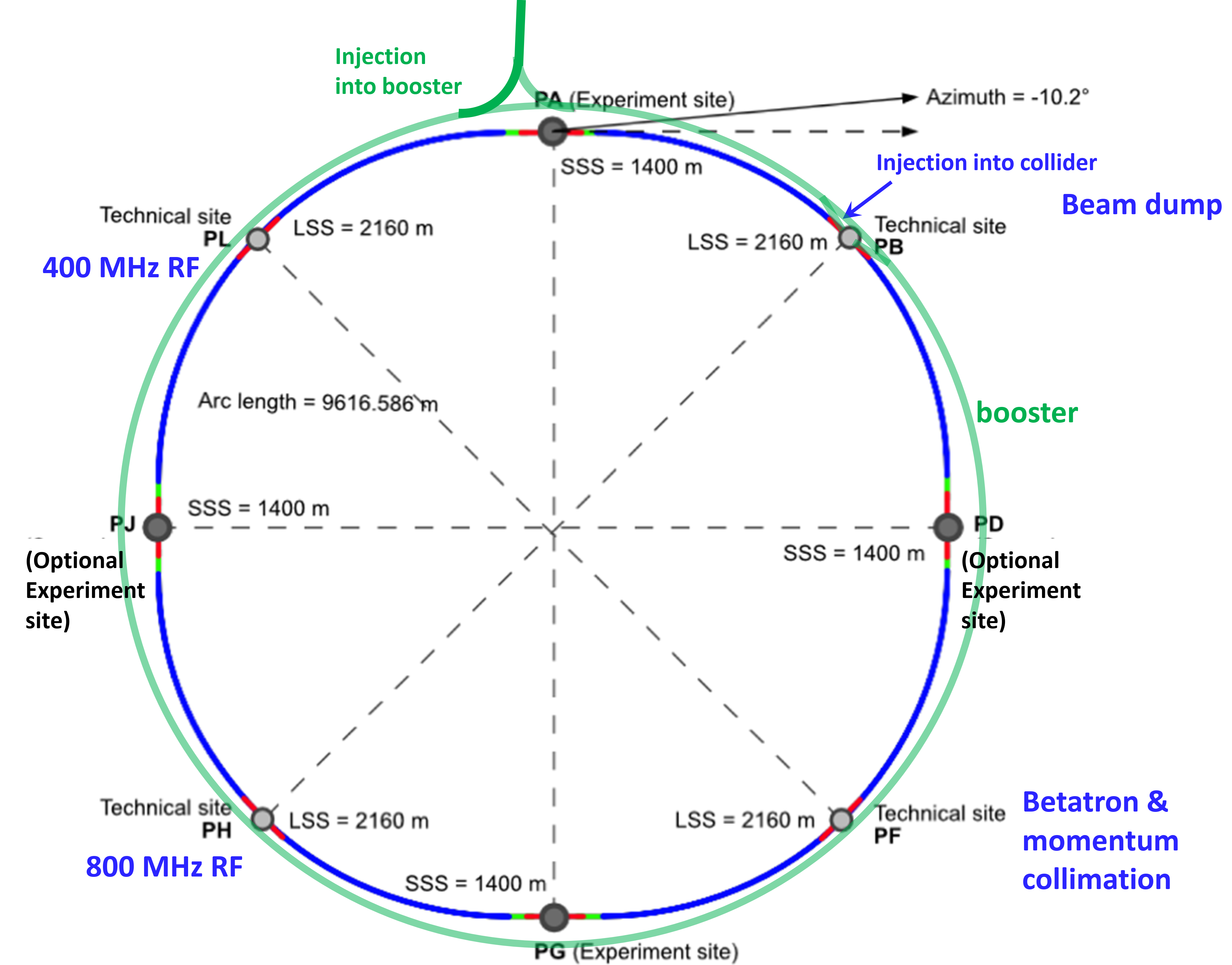}
\includegraphics[width=0.46\linewidth]{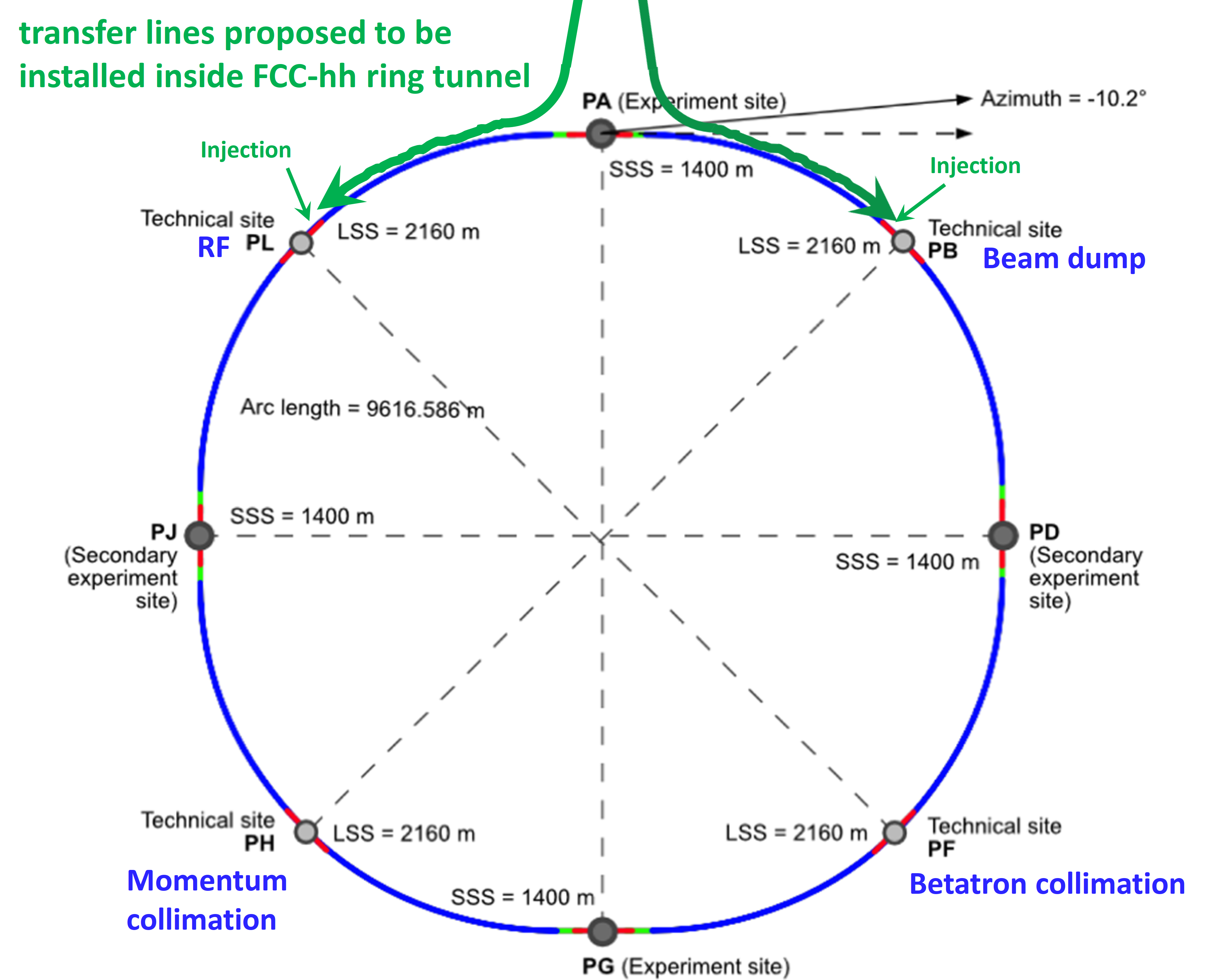}
\vspace*{-3 mm}
\end{center}
\caption{Schematic layout of FCC-ee and its booster (left) and of FCC-hh (right) in the same tunnel infrastructure with a circumference of 91.2 km and strict
four-fold superperiodicity. }  
\label{fig:fcclayout}
\end{figure}

More specifically, the successful 
sequence of LEP and LHC, 
that will stretch over at least 65 years,
has inspired 
the Future Circular Collider (FCC) ``integrated program''\cite{Benedikt:2653673}, based on a three times larger tunnel and set to extend over 70--80 years. 
The latter offers a 
comprehensive long-term plan maximizing physics opportunities. 
Similar to LEP, the first stage is an electron-positron collider,
FCC-ee, that will operate at four different energies, corresponding to the Z pole (at $10^5$ times the LEP luminosity), the W threshold, the Higgs (ZH) production peak, and the ${\rm t}\bar{\rm t}$ threshold, serving 
 as a unique Higgs factory, electroweak \& top factory at highest luminosities.
In a second stage, the hadron collider FCC-hh, with a centre-of-mass energy of about 100 TeV, would provide the natural continuation at energy frontier, with heavy-ion and lepton-hadron collider options.  
FCC-ee and hh cover complementary physics, benefit from 
common civil engineering and 
technical infrastructures (see Fig.~\ref{fig:fcclayout}, 
both building on and reusing CERN’s existing installations. 
In addition, the FCC integrated project allows for a seamless continuation of High Energy Physics after the  HL-LHC.

\begin{figure}[htbp]
\centering
\includegraphics[width=0.8\linewidth]{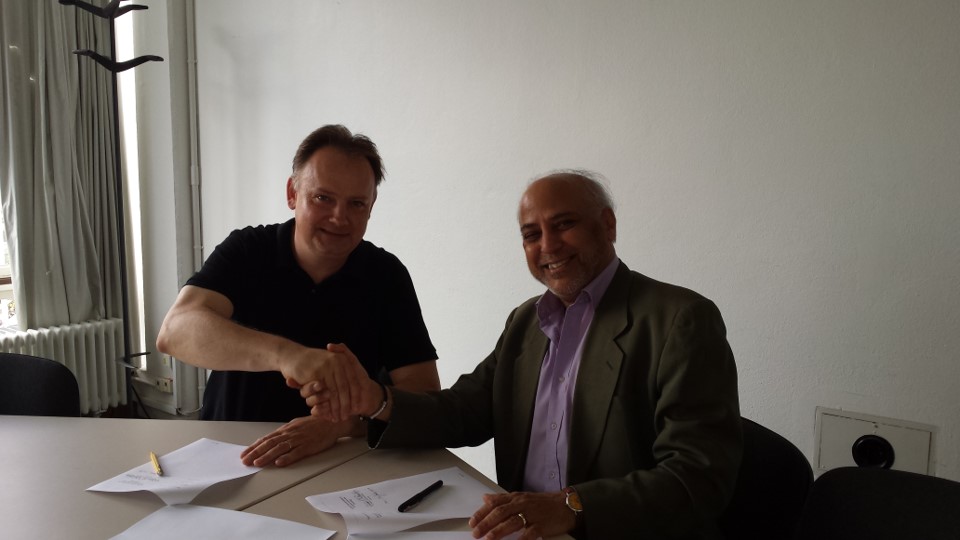}
\caption{In summer 2014, CI Director Swapan Chattopadhyay and CERN's FCC Study Leader Michael Benedikt 
seal CI's membership in the 
international FCC Collaboration.}
\label{fig:fcc}
\end{figure}

\begin{figure}[htb]
\centering
\resizebox{0.85\textwidth}{!}{%
  \includegraphics{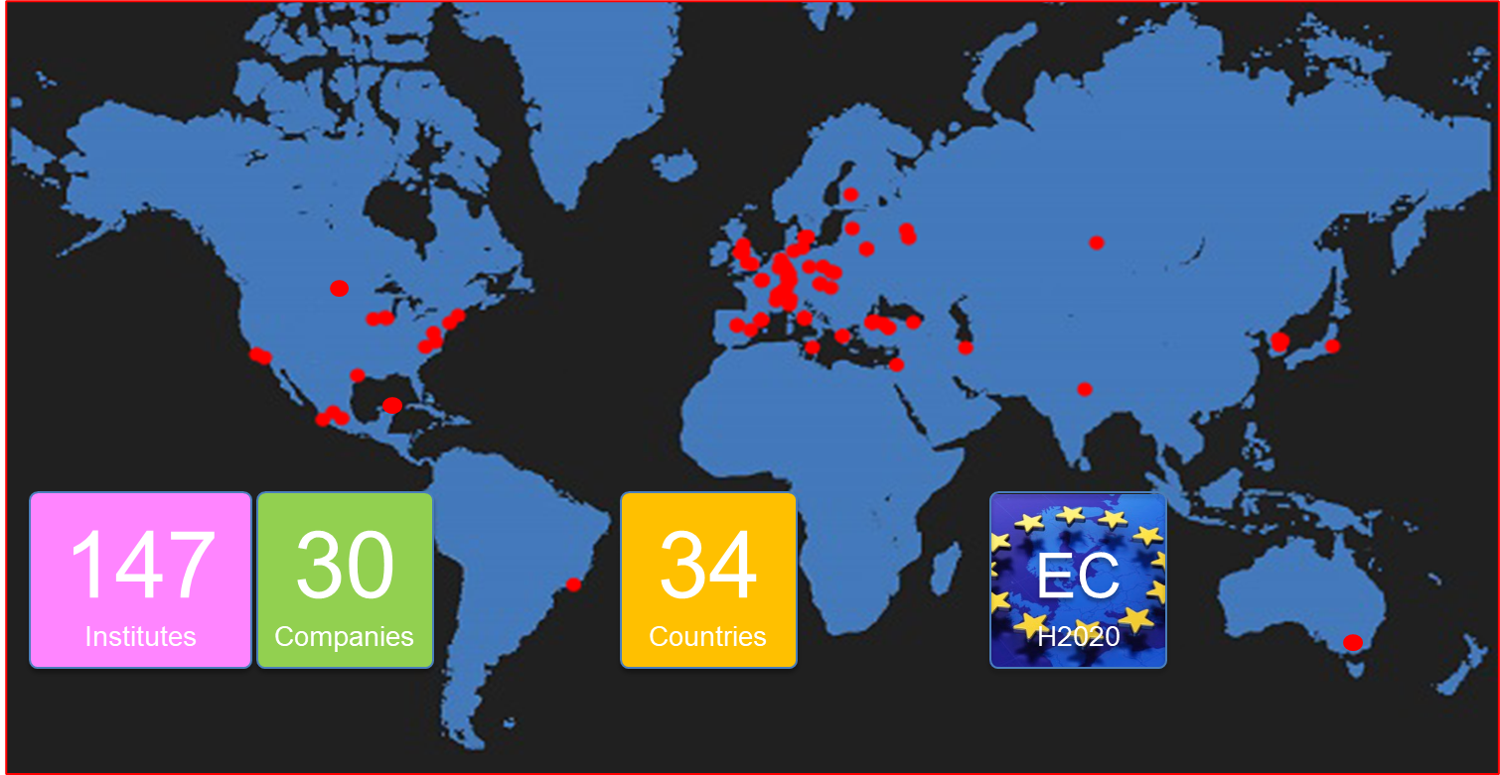}
}
\caption{The FCC collaboration spans the world and has  engaged more than 145 institutions.}
\label{fig:FCCcollab}       
\end{figure}

Swapan Chattopadhyay recognized the FCC merit and potential early on.
Already in summer 2014, shortly after the FCC kick-off event, with the CI he joined
the FCC collaboration;  
indeed, the CI was the first institute 
to enter this new collaboration (Fig.~\ref{fig:fcc}).  
Following the lead of the CI, 
about 150 other institutes from around the world 
equally joined the FCC effort (see Fig.\ref{fig:FCCcollab}).

\section{Electron-Hadron Colliders: EIC and LHeC}
\begin{figure}[htb]
\begin{center}
\includegraphics[width=0.48\linewidth]{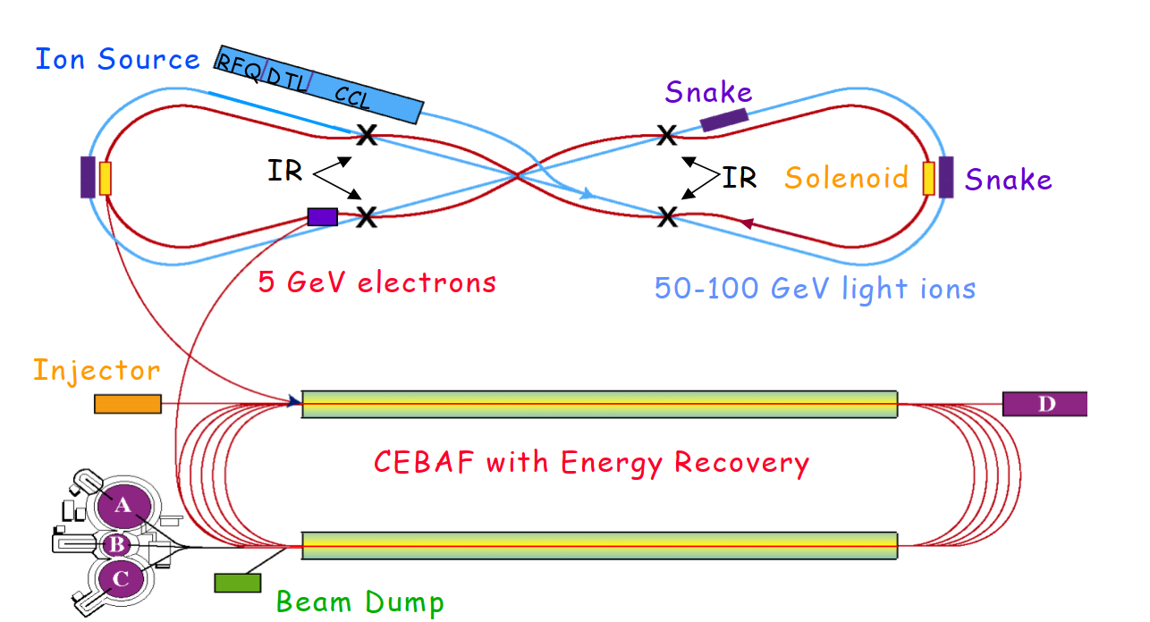}
\includegraphics[width=0.46\linewidth]{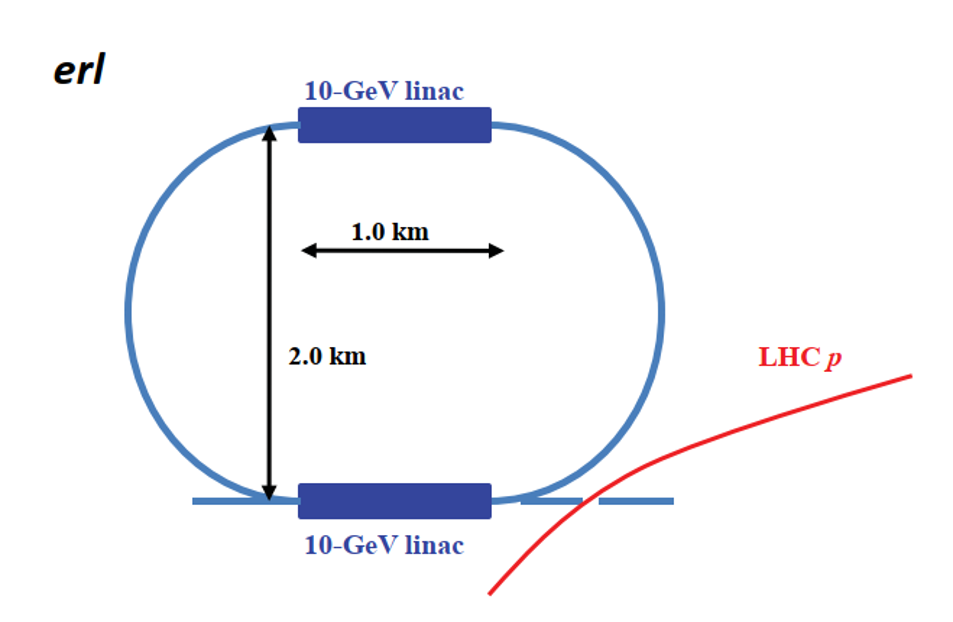}
\vspace*{-3 mm}
\end{center}
\caption{Two ERL-based lepton hadron colliders co-developed by Swapan Chattopadhyay: the JLAB ELIC from 2002\cite{Merminga:584415}, and the ERL-based version of the LHeC first proposed around 2008\cite{zimmermann:1124066,Zimmermann:1269309,Abelleira_Fernandez_2012}.}
\label{fig:lephadcoll}       
\end{figure}

Swapan Chattopadhyay was instrumental in the design of future electron-hadron colliders, such as the ELIC\cite{Merminga:584415} (an earlier, ERL-based version 
of the US Electron Ion Collider, now close to  construction at BNL) and the LHeC at CERN\cite{zimmermann:1124066,Zimmermann:1269309};
see Fig.~\ref{fig:lephadcoll}. 
These linac-ring type colliders based on a recirculating-electron linac can achieve significant luminosity, thanks to energy recovery.
In particular, with his experience from JLAB, Swapan Chattopadhyay brought the concept of energy recovery to the linac-ring LHeC design\cite{zimmermann:1124066,Zimmermann:1269309}.

\section{A Bird's Eye View of Future Colliders}
\begin{figure}[htb]
\begin{center}
\includegraphics[width=0.85\linewidth]{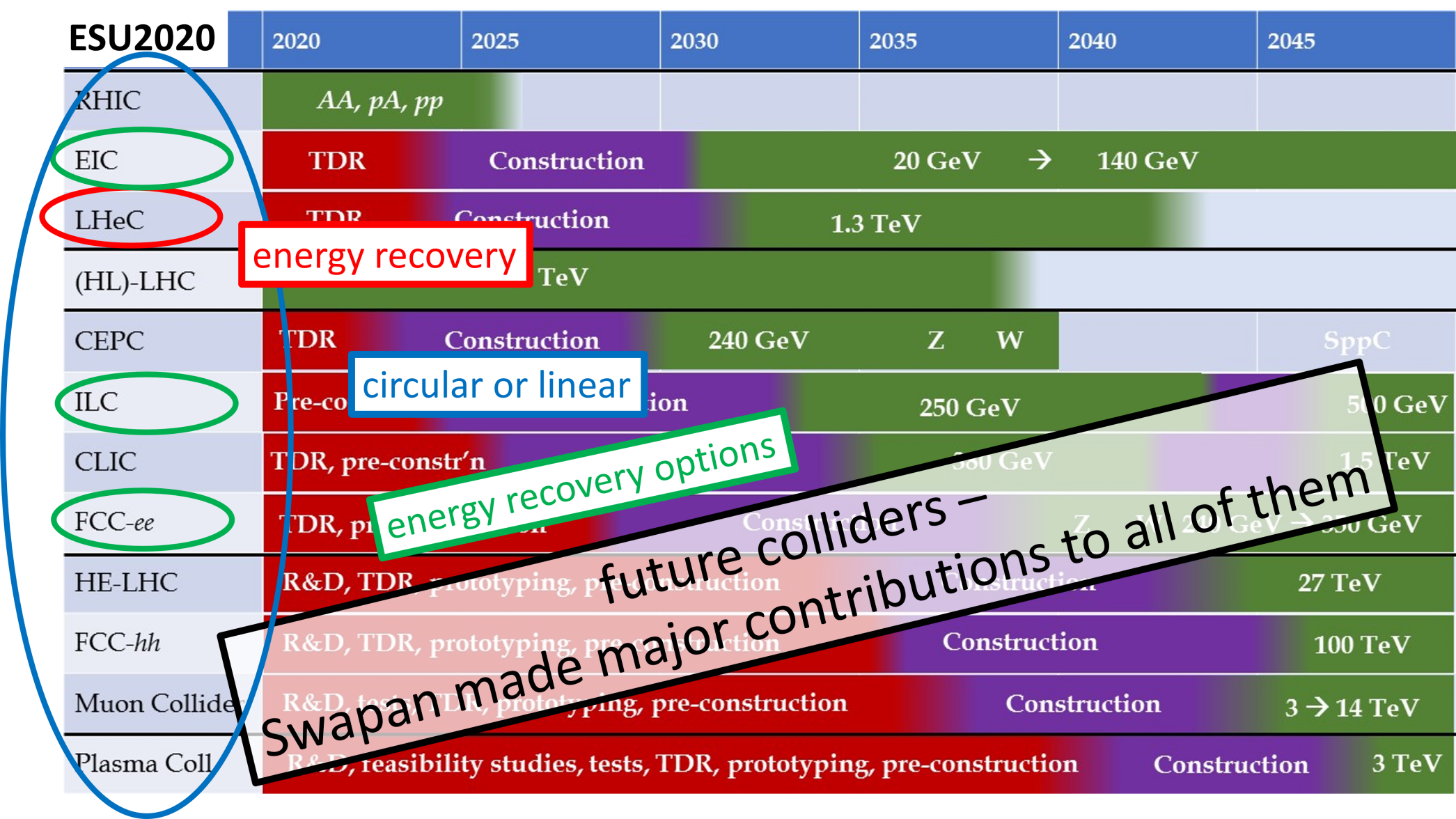}
\vspace*{-3 mm}
\end{center}
\caption{Approximate technically limited timelines of future large colliding-beam facilities\cite{RevModPhys.93.015006}. 
Swapan Chatopadhyay contributed  essential elements 
to all of the various types of colliders, whose operation 
will extend over a century. 
}
\label{fig:futurecoll}       
\end{figure}

In addition to proposed high-energy electron-positron and hadron colliders,
and lepton-hadron colliders, there are other types of future colliders, probably candidates for the next-next generation of colliders, such as 
muon colliders and colliders based on plasma acceleration\cite{RevModPhys.93.015006}.
Figure \ref{fig:futurecoll}, adapted from Ref.~\cite{RevModPhys.93.015006}, 
illustrates  approximate technically limited time lines of future large colliding-beam facilities for the next three 
decades based on the presentations by their proponents given
and briefly discussed at the 2019 European Particle Physics Strategy Update Symposium\cite{preparatorygroup2020physics} and the European Strategy Update 2020 (ESU2020).

It is remarkable that Swapan Chattopadhyay has made essential contributions to all the various types of colliders, thereby laying a solid foundation for the future.

\section{Recover the Energy!}
\begin{figure}[htb]
\begin{center}
\includegraphics[width=0.55\linewidth]{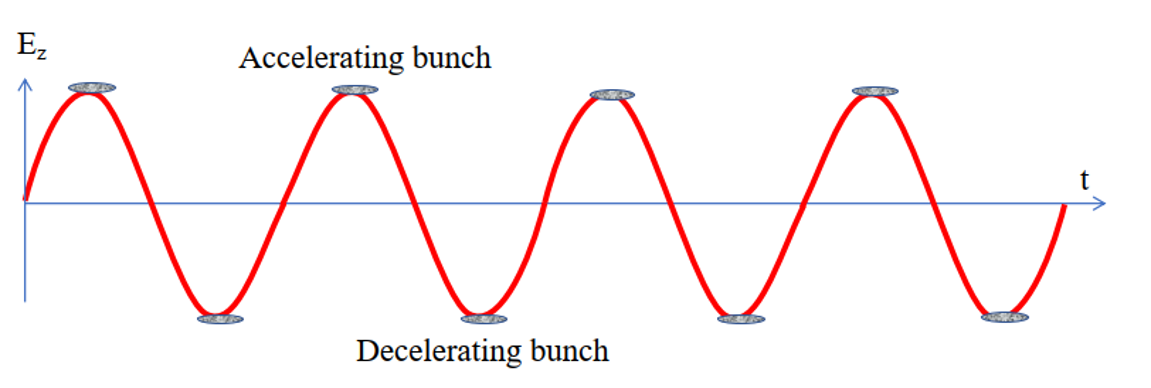}
\includegraphics[width=0.40\linewidth]{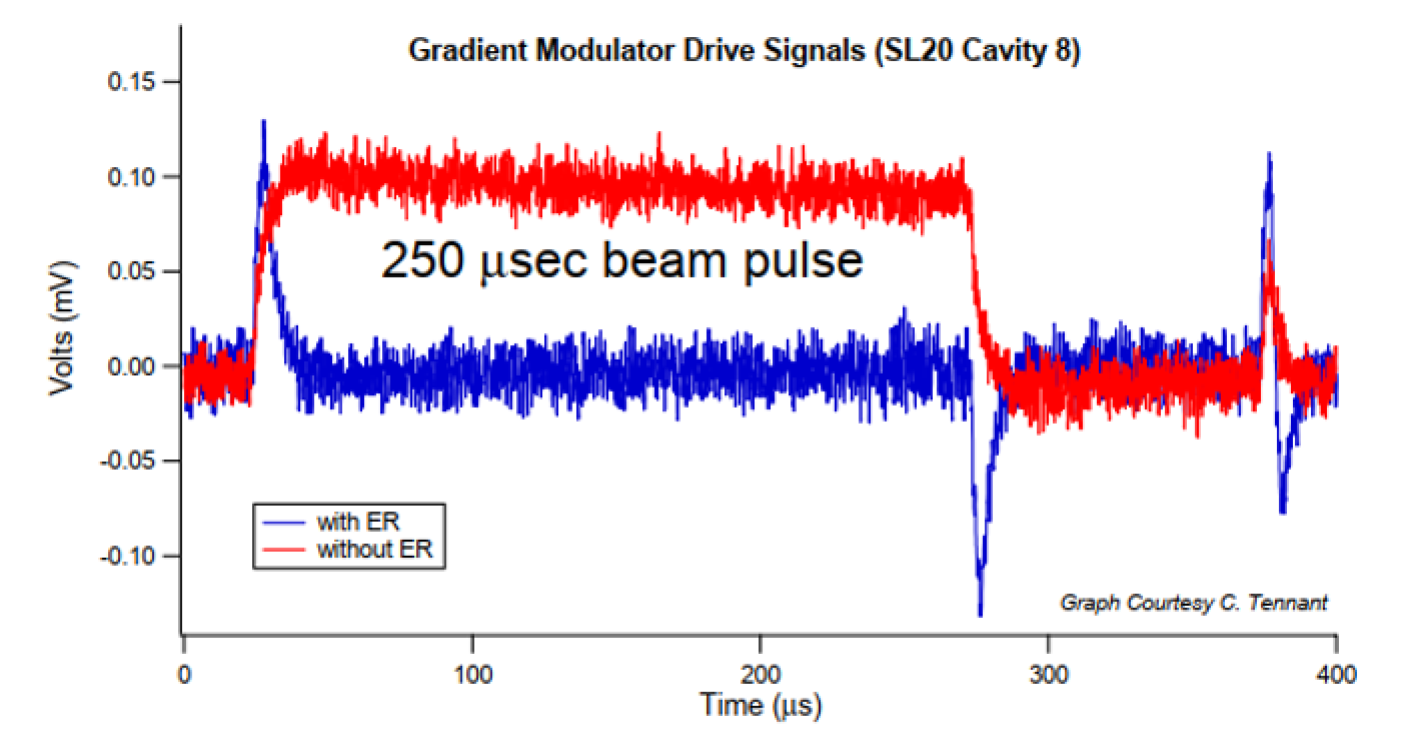}
\vspace*{-3 mm}
\end{center}
\caption{The principle of energy recovery\cite{ERLprinciple} (left);
forward power required by a linac  
cavity with and without energy recovery as 
measured at the CEBAF experiment\cite{Bogacz:2003zz} (right).
}
\label{fig:ERLprinciple}       
\end{figure}

The principle of energy recovery is shown 
in Fig.~\ref{fig:ERLprinciple}. 
Around 2002/2003 Swapan Chattopadhyay 
oversaw
a pioneering experiment on the 
recirculating linear accelerator 
successfully, which demonstrated GeV scale energy recovery with a high ratio of accelerated-to-recovered energies (50:1) 
\cite{Bogacz:2003zz}.

\begin{figure}[htb]
\begin{center}
\includegraphics[width=0.45\linewidth]{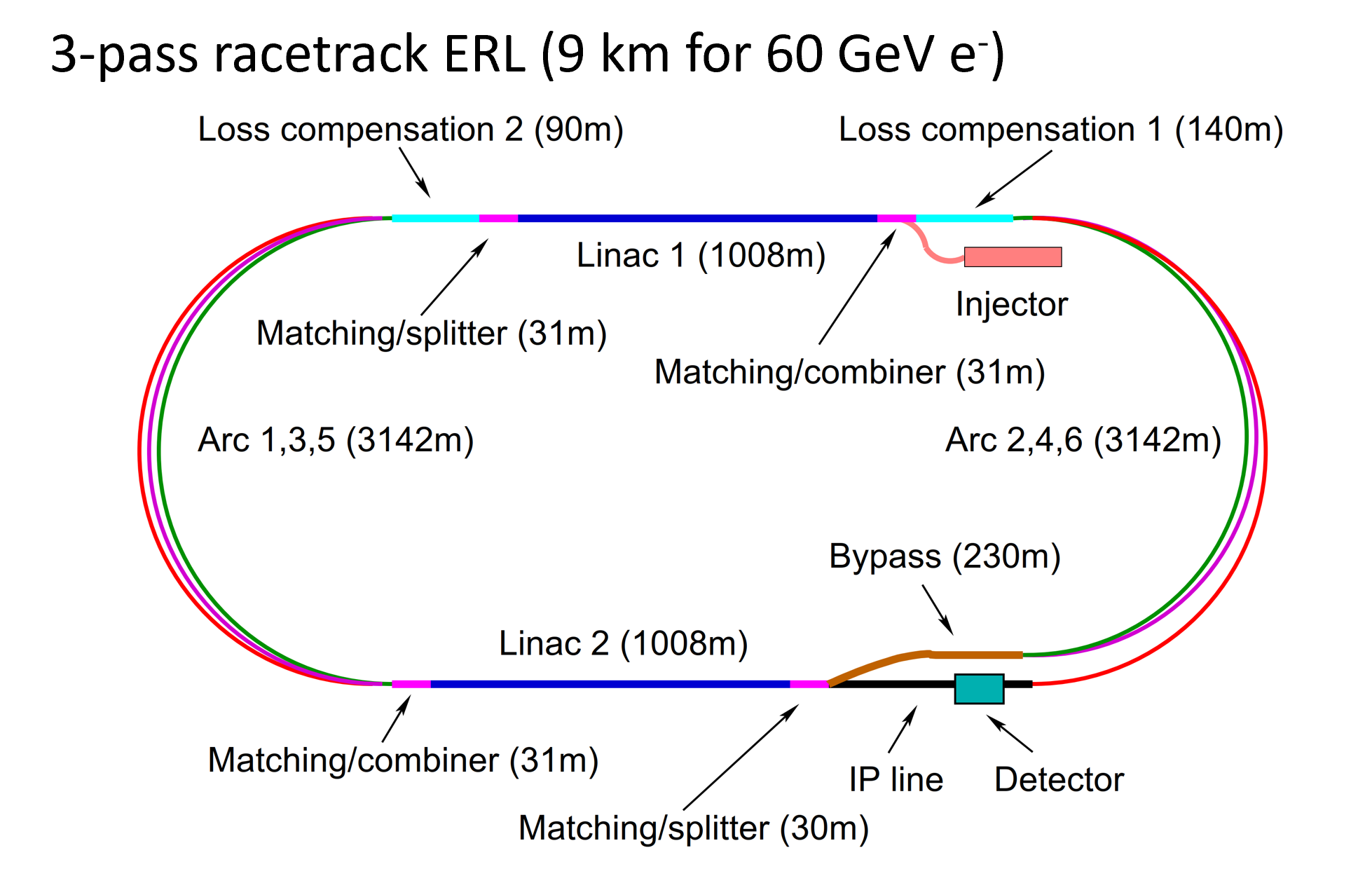}
\includegraphics[width=0.45\linewidth]{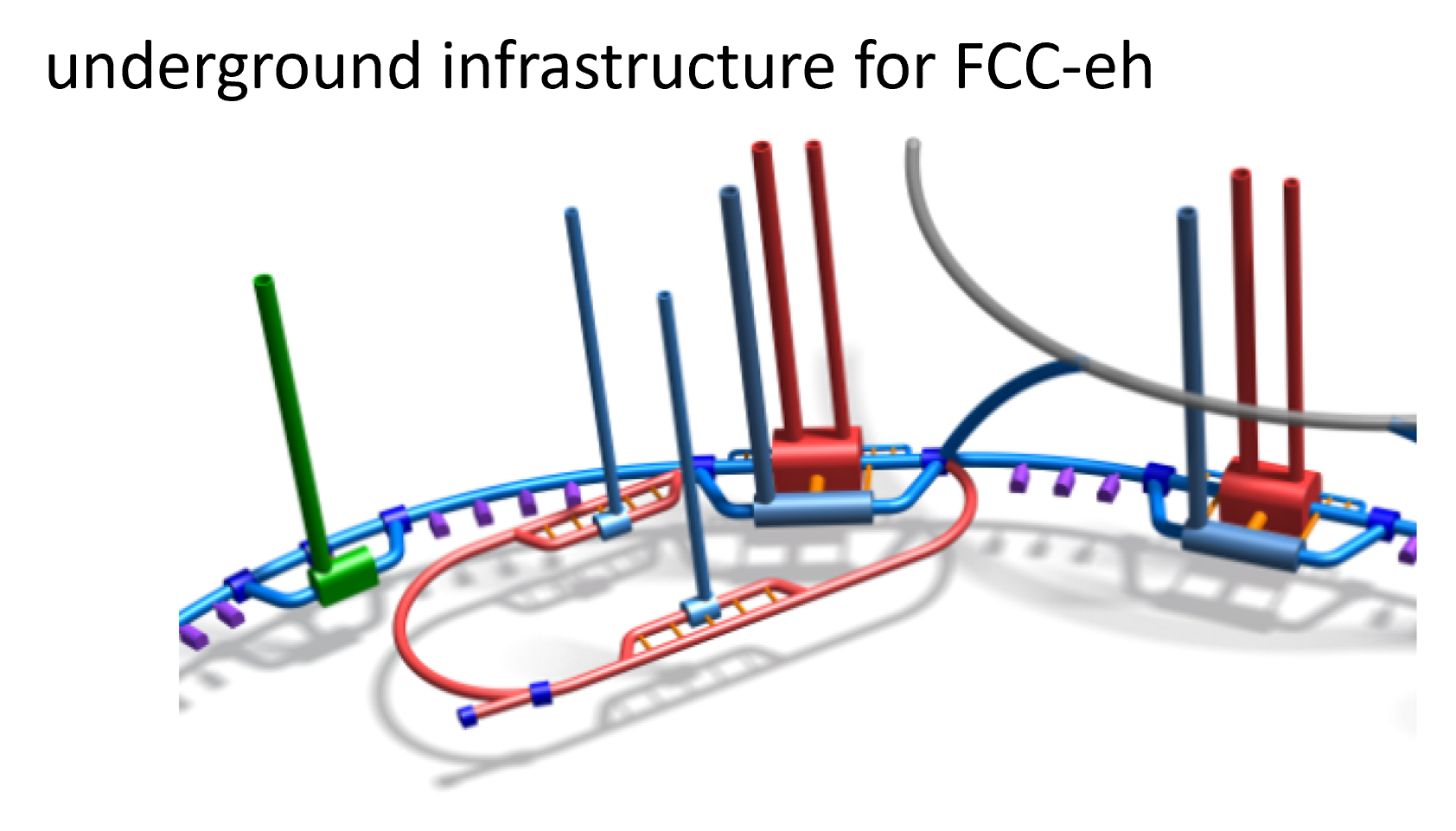}
\vspace*{-3 mm}
\end{center}
\caption{Sketch of the 3-pass ERL layout for LHeC\cite{Abelleira_Fernandez_2012} (left), and of the underground 
infrastructure for FCC-eh 
\cite{fcchh} (right).
}
\label{fig:ERL-ep-coll}       
\end{figure}

\begin{figure}[htb]
\begin{center}
\includegraphics[width=0.85\linewidth]{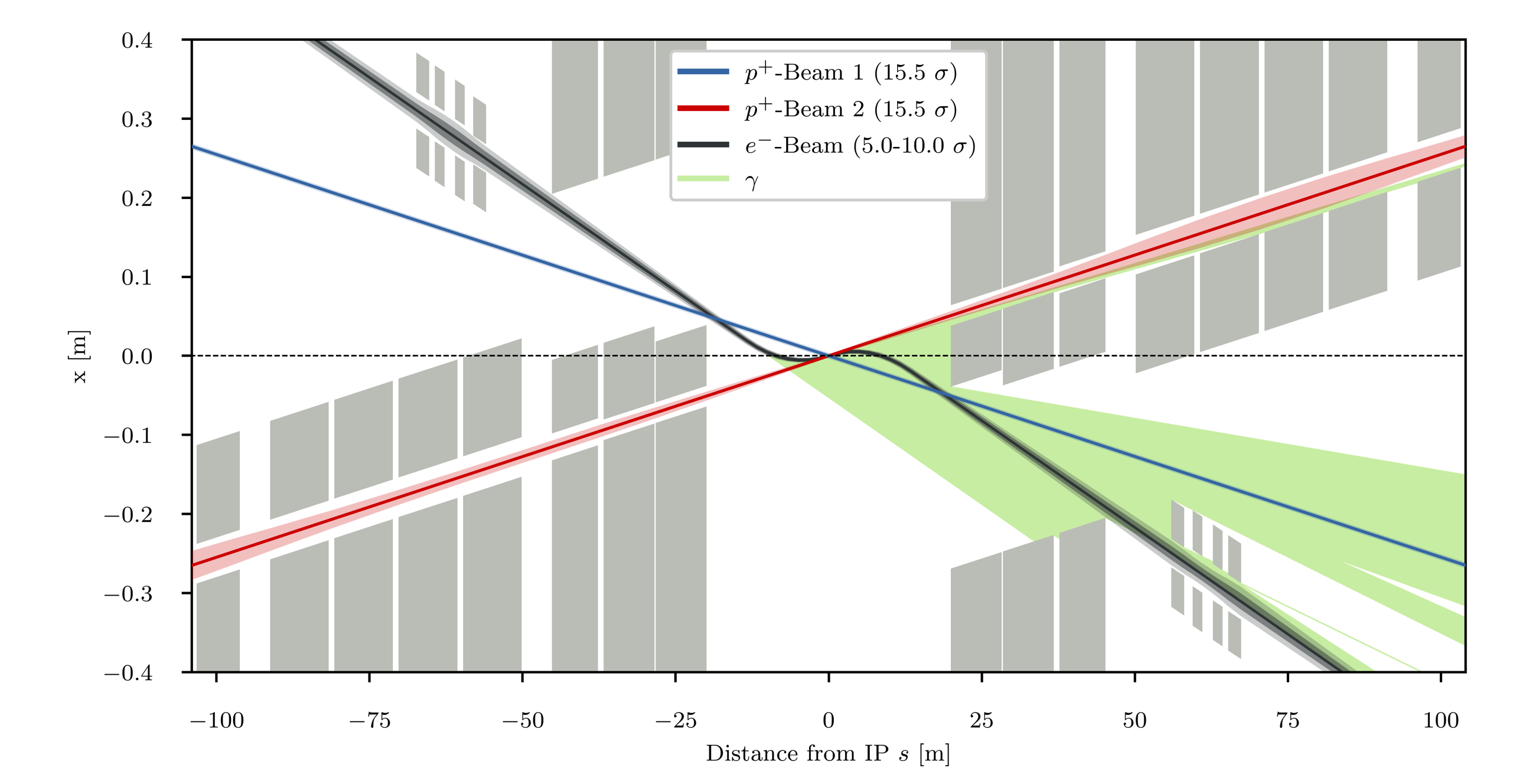}

\vspace*{-3 mm}
\end{center}
\caption{Sketch of the 3-beam
interaction region for FCC-eh\cite{fcchh}.}
\label{fig:3beamip}       
\end{figure}

Proposed future ERL-based lepton-hadron colliders at CERN include the LHeC and FCC-eh, 
where 50--60 GeV electrons from a racetrack-shape multi-turn ERL 
are collided with the 
7 or 50 TeV protons of the LHC or 
future circular hadron collider (FCC-hh), respectively.
These are illustrated in Fig.~\ref{fig:ERL-ep-coll}.
Aside from the ERL design, another
common challenge for LHeC and FCC-eh
is the interaction region, which must accommodate the two 
counterpropagating
proton beams (colliding elsewhere around the LHC or FCC-hh rings),
and the electron beam from the ERL.
Figure \ref{fig:3beamip} shows an example configuration for FCC-eh.

\begin{figure}[htb]
\begin{center}
\includegraphics[width=0.60\linewidth]{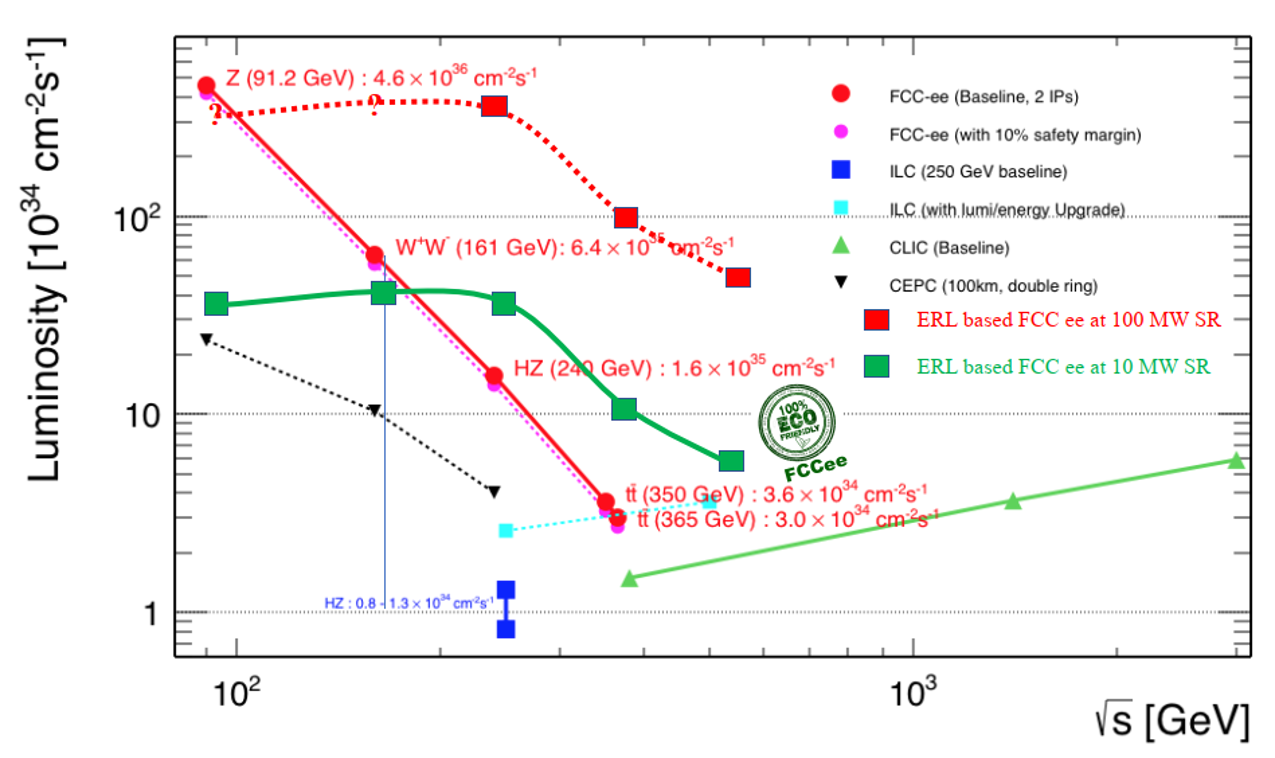}
\includegraphics[width=0.36\linewidth]{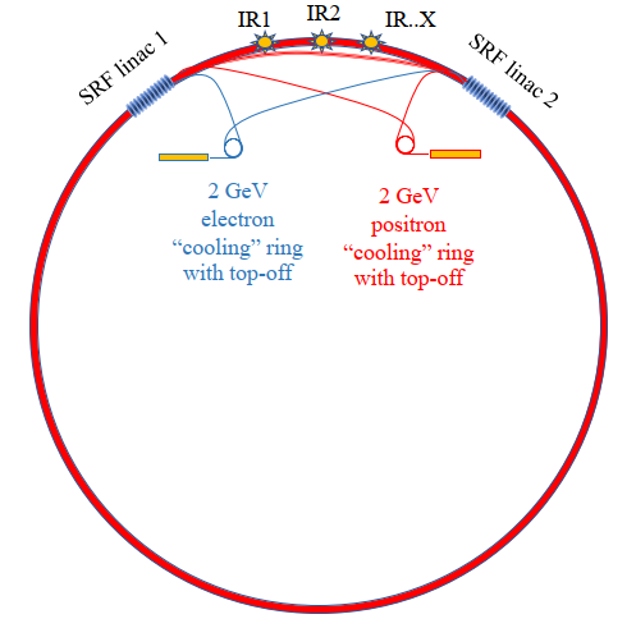}
\vspace*{-3 mm}
\end{center}
\caption{Luminosities for various options for high-energy e$^+$e$^-$ collider \protect\cite{LITVINENKO2020135394,ERLprinciple} (left), 
and sketch of a possible layout of an ERL-based circular 
e$^+$e$^-$ collider with linacs separated by 1/6th of the 100 km circumference \protect\cite{LITVINENKO2020135394} (right).}
\label{fig:fcceeERL}       
\end{figure}

ERL-based variants  
are also proposed for e$^+$e$^-$ colliders. 
An ERL-based upgrade option, or alternative, for the circular  FCC-ee (see Fig.~\ref{fig:fcceeERL})  
promises higher luminosity and energy reach.
ERL variants, like LERC\cite{telnov2021} and RELIC\cite{lit2022}, are also being 
advocated for linear colliders. 
This would be in line with history as the very first proposals of linear colliders in the 
1960s\cite{Tigner:1965wf} and 1970s\cite{Amaldi:1975hi,Gerke:1979zs}
were all based on energy recovery,
and as, conversely, also the concept of 
energy recovery was first proposed with a linear  collider in mind.
For ERL-based linear colliders the accelerating
and decelerating bunches should not collide.
This could be achieved, for example, 
by introducing electrostatic separators
\cite{PhysRevSTAB.8.010701}
or by use of a dual-axis linac\cite{PhysRevAccelBeams.19.083502}.

\begin{figure}[htb]
\begin{center}
\includegraphics[width=0.85\linewidth]{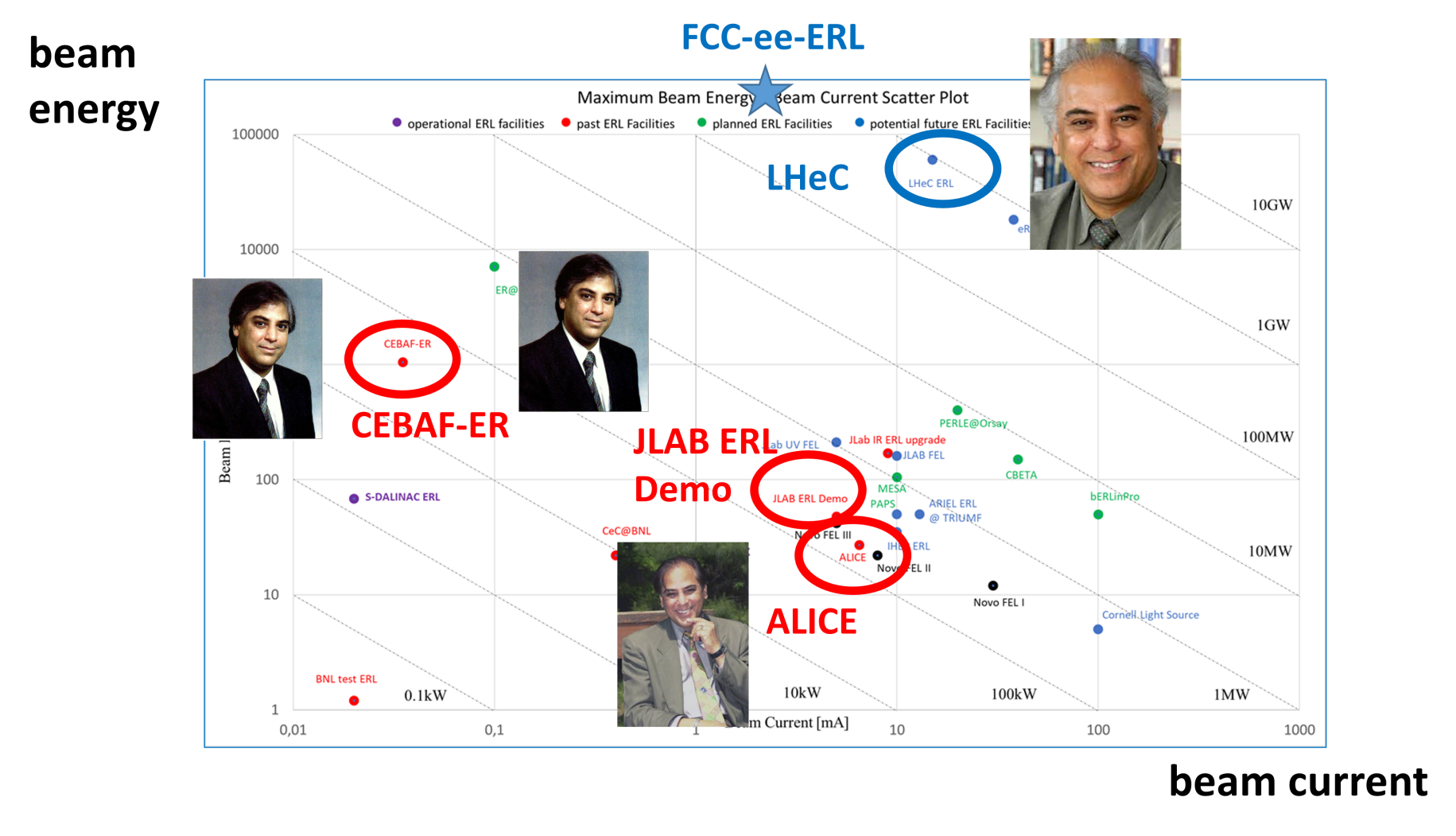}
\vspace*{-3 mm}
\end{center}
\caption{Beam energy 
as a function of beam current for  various past, operating, planned, and proposed ERLs  (modified from F.~Hug et al.\protect\cite{ariesms28})
with photos of Swapan Chattopadhyay superimposed.}
\label{fig:erlls}       
\end{figure}

Figure \ref{fig:erlls} illustrates
the ERL landscape on a double-logarithmic scale and 
highlights the various ERL projects,
to which Swapan Chattopadhyay made
significant contributions in the various phases of his long career.

\section{Neutrino Factories and Muons Colliders}
At the same beam energy, muons bent in a circle lose $1.6\times 10^9$ times less energy from synchrotron radiation than electrons or positrons. Also the beamstrahlung emitted during the collision is dramatically reduced. 
Swapan Chattpadhyay took an early interest in muon colliders and analyzed the 
critical issues\cite{Chattopadhyay:1994td}.

Though muons radiate less, there are other challenges.
The muons are unstable and decay within 
a few 100s to 1000s of turns.
This requires rapid acceleration --- 
perhaps plasma acceleration could be an option,
since gradients are extremely high, and the muons typically of rather low intensity.

Another issue is the neutrino radiation hazard 
caused by the muon decay\cite{King:1999rr}, which may 
limit the maximum muon energy attainable on earth 
to about 10 TeV or at most a few 10s of TeV.
The cross section of neutrinos interacting with matter increases linearly with energy, and 
the maximum neutrino flux roughly with the square of the energy due to the Lorentz boost.

Several production schemes for muons are proposed.
The first is proton-beam driven: 
Protons hitting a target generate pions, which decay into muons.
In this scheme, developed by the US-MAP collaboration\cite{Delahaye:2014vvd}, the muon beams are generated with large 
emittance. For a collider the muon beam must, therefore,  be cooled, and its 6D emittance be reduced --- and rapidly --- by six to eight orders of magnitude. The innovative technique of 
ionization cooling is proposed for this purpose.
Ionization cooling was first demonstrated by the UK's MICE experiment.

Another, more recently suggested 
production scheme, called LEMMA\cite{Antonelli:2015nla,PhysRevAccelBeams.23.051001}, 
is based on positrons
at an energy of about 45 GeV, which annihilate with electrons at rest into muon pairs.
For reasons of energy efficiency, 
this requires a large 45 GeV e$^+$ ring, like the full-energy booster of the FCC-ee,
and offers a possible upgrade path to FCC-$\mu\mu$
\cite{Zimmermann:2018wfu,Zimmermann-ipac22},
which becomes  most powerful if combined with the Gamma Factory concept\cite{Krasny:2015ffb} to realize a highly-intense  
positron source. 

In 1993, Swapan Chattopadhyay and co-workers considered yet another scheme of 
muon generation, namely 
photoproduction from a primary 60 GeV electron beam
hitting a target, based on a proposal by  
W.A.~Barletta and A.M.~Sessler\cite{Barletta:1993rq}.

\begin{figure}[htb]
\begin{center}
\includegraphics[width=0.85\linewidth]{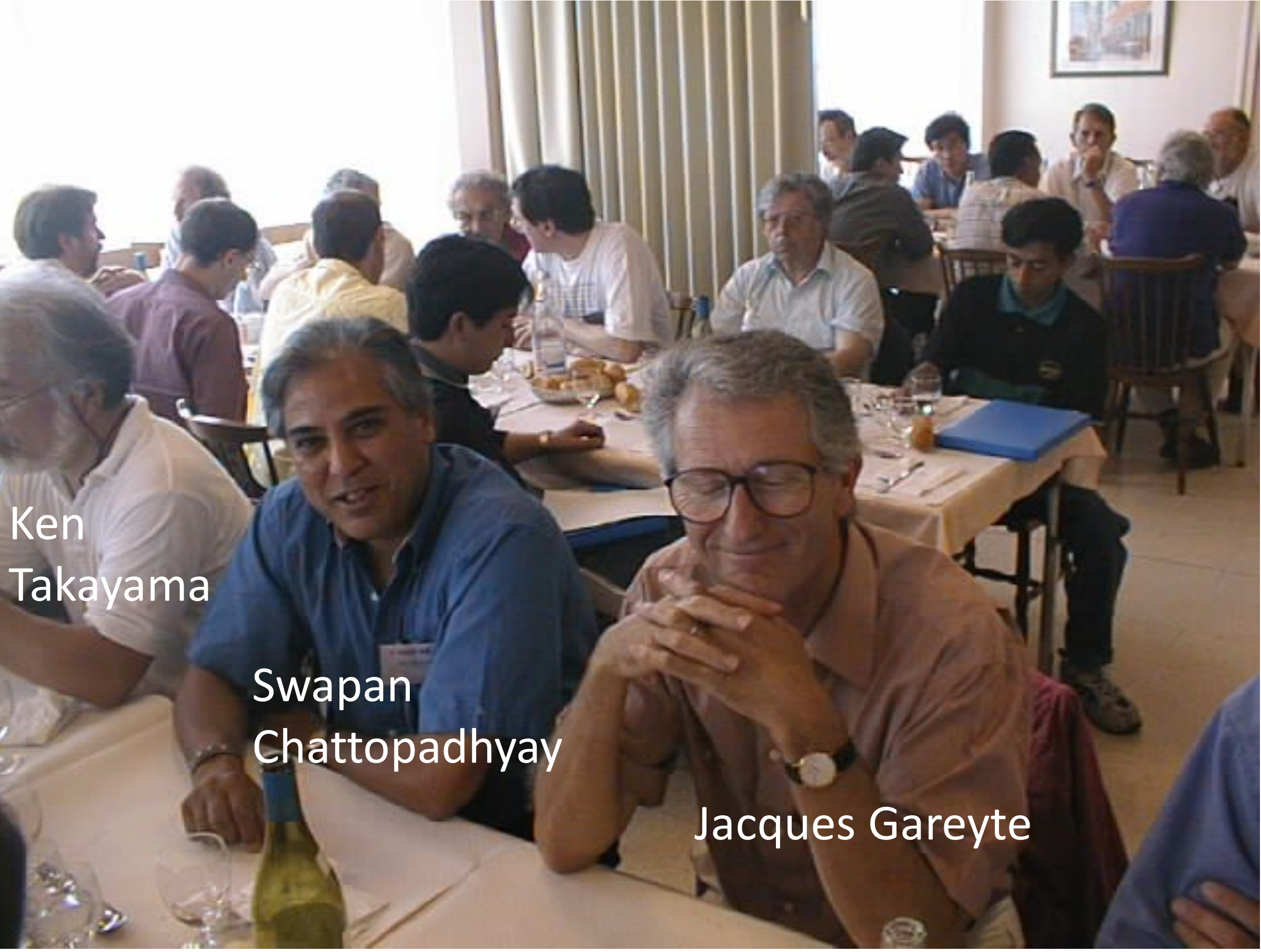}
\vspace*{-3 mm}
\end{center}
\caption{Ken Takayama, Swapan Chattopadhyay and Jacques Gareyte at $\nu$Fact99 in Lyon (Courtesy CERN).}
\label{fig:lyon}       
\end{figure}

An intermediate step towards a muon collider 
could be a neutrino factory, e.g.\cite{Palmer:410060,Boscolo:2018ytm}.  
Figure \ref{fig:lyon} shows a photograph from the $\nu$Fact'99 workshop lunch in Lyon, with 
Swapan Chattopadhyay surrounded by  
KEK and CERN experts.

\section{Plasma and Crystal Colliders}
\begin{figure}[htb]
\begin{center}
\includegraphics[width=0.85\linewidth]{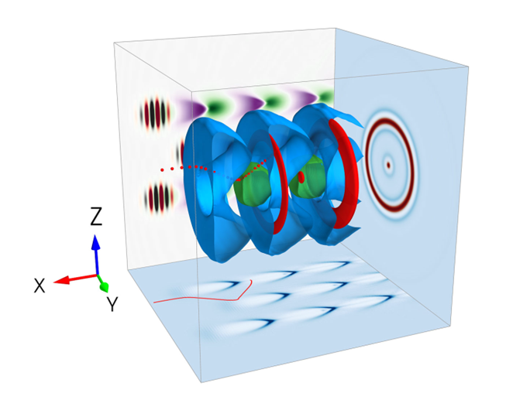}
\vspace*{-3 mm}
\end{center}
\caption{Schematic of positron ballistic injection scheme with two lasers \protect\cite{Xu:2019zov}. The blue and green colors are contour surfaces of electron densities for the ``donut'' and center ``bubbles'', respectively. The red color represents injected positrons. For more details see Ref.~\protect\cite{Xu:2019zov}.}
\label{fig:posaccel}       
\end{figure}

Higher accelerating gradients than in conventional accelerators can be sustained in plasmas.
Accelerating plasma waves can be excited either
by a high-energy 
charged particle beam (beam-driven plasma wake field acceleration --- PWFA) 
or by a high-power laser (laser-driven plasma wake field acceleration --- LWFA). 
Detailed scenarios have been developed for 
electron-positron colliders based on either
PWFA\cite{adli2013beam,chen2021mathrmemathrme}
or LWFA\cite{Leemans:2009zza,Schroeder:2016mrg}. 
Key parameters are rather similar for the two approaches, with 
plasma electron densities between $2\times 10^{16}$~cm$^{-3}$  and 
$10^{17}$~cm$^{-3}$, energy gains per stage between 
5 and 25 GeV, and geometric gradients of 1 to 2.3 
GV/m. 
For e$^+$e$^-$ collider applications, 
it is necessary not only to accelerate electrons but also positrons,
and to do so without unacceptable beam quality degradation.  
This is a major outstanding question and an active area of research. As a possible solution, for the case of LWFA, more complex schemes with multiple driving laser are being developed. An example is shown in Fig.~\ref{fig:posaccel}\cite{Xu:2019zov}.

Even much higher gradients still are possible in crystals. 
The maximum field is given by\cite{Shiltsev:2012zzc} 
\begin{equation}
E_{0} \approx \frac{m_{e}c\omega_p}{e}
\approx 100 \left[ \frac{\rm GeV}{\rm m} \right]
\sqrt{ n_{0} [10^{18}~{\rm cm}^{-3}]}\; ,
\end{equation}
with $\omega_{p}$ the angular 
plasma frequency and $n_{0}$ the electron density.
With electron densities of order $n_{0}\approx 10^{22}$~cm$^{-3}$
to $5\times 10^{24}$~cm$^{-3}$
in a crystal, peak gradients of 10--1000 TV/m would be within reach.

Accelerating fields in a crystal waves could be excited by drivers with adequate wavelength, that is 
not by conventional lasers, but rather by X-ray lasers
\cite{Shiltsev:2012zzc}.

\begin{figure}[htb]
\begin{center}
\includegraphics[width=0.85\linewidth]{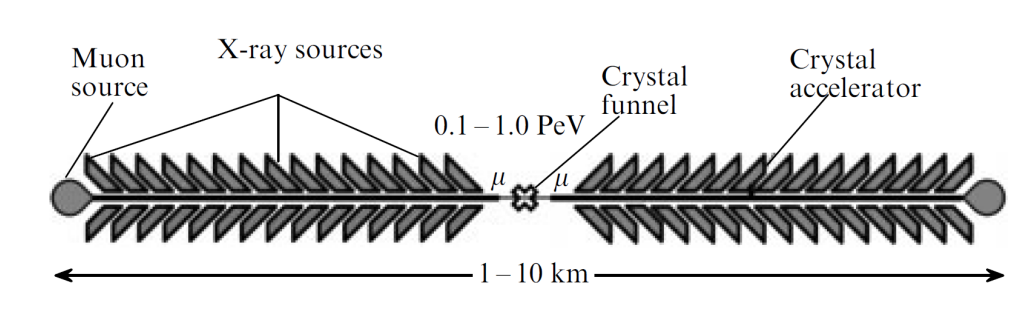}
\vspace*{-3 mm}
\end{center}
\caption{Concept of a linear X-ray crystal muon collider 
(V.~Shiltsev)\cite{Shiltsev:2019clx}.}
\label{fig:cc}       
\end{figure}

The recently developed thin film compression technique
\cite{hakimi} provides an economic path to generating single cycle coherent X-ray pulses and, thereby, to TV/cm acceleration at solid state densities. 
The concept of a  far future X-ray driven crystal collider is illustrated in Fig.~\ref{fig:cc}.

\section{Towards the Skies}
\begin{figure}[htb]
\begin{center}
\includegraphics[width=0.85\linewidth]{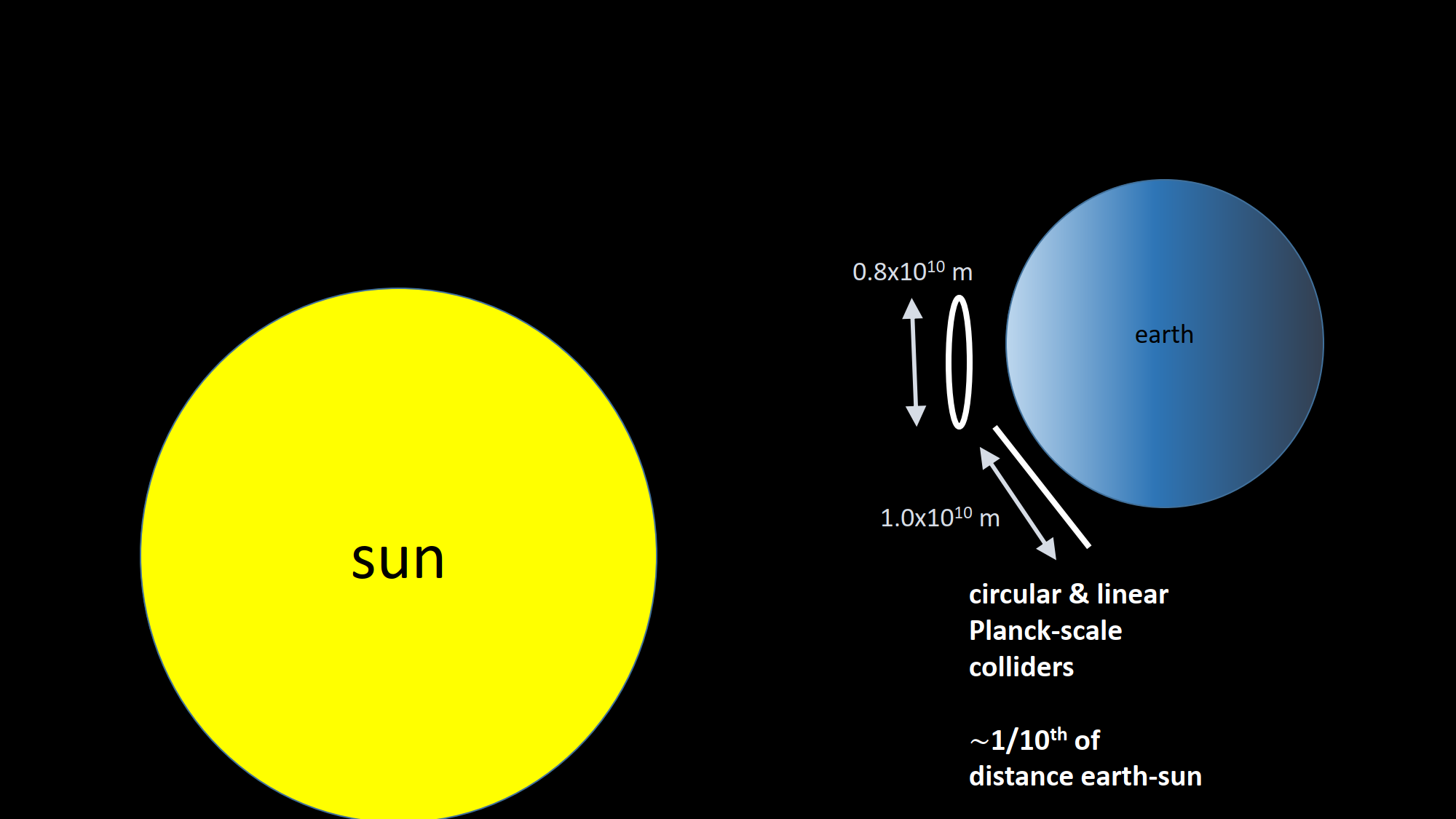}
\vspace*{-3 mm}
\end{center}
\caption{Circular and linear Planck scale colliders operating at the Schwinger critical field 
comfortably fit into the solar system\cite{Chen:1997ai,ZIMMERMANN201833}.}
\label{fig:planck}       
\end{figure}

The ultimate limit on electromagnetic acceleration
in vacuum is given by the 
Schwinger critical field $E_{\rm cr}\approx  10^{12}$~MV/m, or equivalently $B_{\rm cr}\approx 4.4 \times 10^9$~T, at which the QED vacuum breaks down.
To reach the Planck scale of 10$^{28}$~eV,
linear or circular colliders would need to have a size of order $10^{10}$~m, if operated close to the critical field\cite{Chen:1997ai,ZIMMERMANN201833}.
Such colliders are illustrated in Fig.~\ref{fig:planck}. 
This prospect was examined already 
in the 1990s and judged to be “not an inconceivable task for an  advanced technological society”
\cite{Chen:1997ai}.
Following the FCC a possible next or next-next step in this direction could be a circular collider on the moo

\section{Epilogue}
\begin{figure}[htb]
\begin{center}
\includegraphics[width=0.45\linewidth]{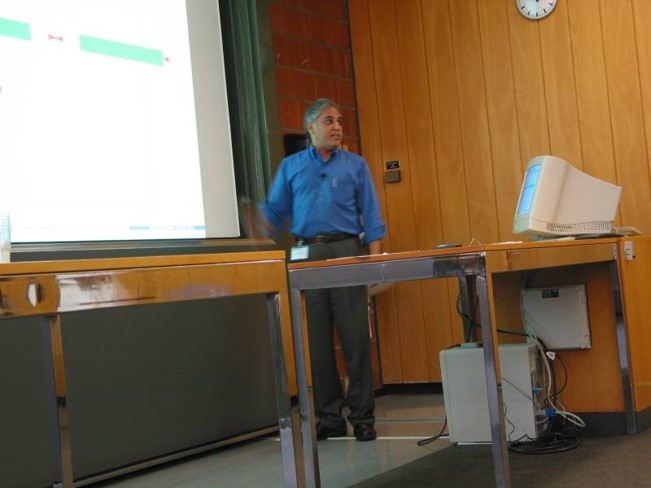}
\includegraphics[width=0.45\linewidth]{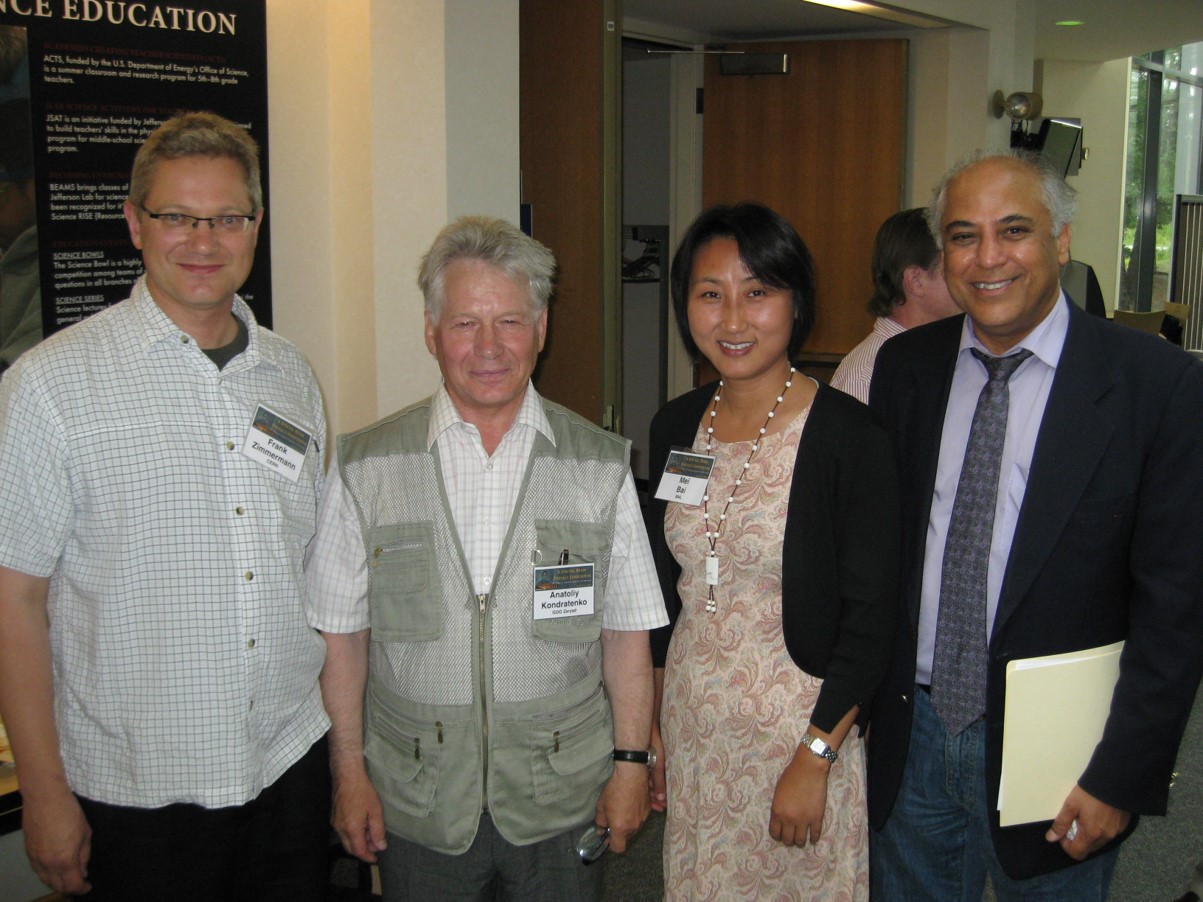}
\vspace*{-3 mm}
\end{center}
\caption{Swapan Chattpadhyay lecturing at the ICFA Nanobeam workshop in Kyoto, Japan, 2005 (left);
Frank Zimmermann, Anatoly Kondratenko, Mei Bai, and 
Swapan Chattopadhyay during the 70th anniversary symposium
for Slava Derbenev, in Newport News, 2010 (right).
}
\label{fig:Swapan0510}       
\end{figure}

During the past 40 years, the collider progress was stunning,
as could be testified by the participants of ICFA Nanobeam workshop 2005 and 
the Slava Derbenev Symposium 2010 
(see Fig.~\ref{fig:Swapan0510}).

Thanks to eminent colleagues like Swapan Chattpadhyay,
we are also well prepared for times ahead --- 
thank you, Swapan, and my warmest wishes for the future!

\section*{Acknowledgments}
I would like to thank A.~Valishev, and L.~Lopez from< FNAL,  
and their colleagues, 
for organizing the symposium 
celebrating S.~Chattopadhyay’s retirement on 30 April 2021.

This work was supported, in parts, by funding from the European Union’s Horizon 2020 Research and Innovation programme under Grant Agreement No 101004730 (iFAST).


\bibliographystyle{ws-ijmpa}
\bibliography{SwapanProc}

\end{document}